\documentclass[11pt]{article}

\usepackage{amssymb,amsmath,amsthm}	
\usepackage{a4wide}
\usepackage[utf8x]{luainputenc}               	
\usepackage{graphicx}      										
\usepackage{tikz}          										
\usepackage{dsfont}
\usepackage{hyperref}
\usepackage{multirow}
\usepackage{subcaption}
\usepackage{amsmath}
\usepackage{xspace}
\usepackage{cleveref}
\usepackage{placeins}
\usepackage{listings}
\newtheorem{definition}{Definition}[]

\newtheorem{remark}{Remark}[]

\newtheorem{example}{Example}[]

\newcommand{\R}{\ensuremath{\mathbb{R}}}

\newcommand{\N}{\ensuremath{\mathbb{N}}}

\DeclareMathOperator{\Unifd}{Unif_{d}}
\DeclareMathOperator{\Unifc}{Unif_{c}}

\begin{document}
	

	\title{SABCEMM-\\
	A Simulator for \\
	Agent-Based Computational Economic Market Models }
	
	\author{Torsten Trimborn\footnote{IGPM, RWTH Aachen, Schinkelstrasse 2, 52056 Aachen, Germany} \footnote{Corresponding author: trimborn@igpm.rwth-aachen.de}, Philipp Otte\footnote{MathCCES, RWTH Aachen, 52056 Aachen, Germany}, Simon Cramer\footnote{RWTH Aachen, 52056 Aachen, Germany},\\ Maximilian Beikirch\footnotemark[4], Emma Pabich\footnotemark[4], Martin Frank\footnote{Karlsruhe Institute of Technology, Steinbuch Center for Computing, Hermann-von-Helmholtz-Platz 1, 76344 Eggenstein-Leopoldshafen, Germany}}
	
\maketitle




\begin{abstract}
We introduce the simulation tool \emph{SABCEMM} (Simulator for Agent-Based Computational Economic Market Models) for agent-based computational economic market (ABCEM) models.
Our simulation tool is implemented in C++ and we can easily run ABCEM models with  several million agents.
The object-oriented software design enables the isolated implementation of building blocks for ABCEM models, such as agent types and market mechanisms.
The user can design and compare ABCEM models in a unified environment by recombining existing building blocks using the XML-based SABCEMM configuration file.
We introduce an abstract ABCEM model class which our simulation tool is built upon.
Furthermore, we present the software architecture as well as computational aspects of SABCEMM. 
Here, we focus on the efficiency of SABCEMM with respect to the run time of our simulations.
 We show the great impact of different random number generators on the run time of ABCEM models.
The code and documentation is published on GitHub at \url{https://github.com/SABCEMM/SABCEMM}, such  that all results can be reproduced by the reader.
 \end{abstract}

\section{Introduction}
Over the last two decades, the new research field of \textit{Econophysics} benefited from a rapidly increasing community and gained lots of momentum \cite{bouchaud2002introduction}.
 Due to several financial crises, the interest in new financial market models has risen, not only in the scientific society \cite{farmer2009economy, bouchaud2008economics} but especially in the work of practitioners like Trichet \cite{Trichet2010} and Bernanke \cite{Bernanke2010}.
One subject of \textit{Econophysics} are so called agent-based computational economic market (ABCEM) models. These models resemble an artificial market of interacting agents usually analyzed with the help of Monte Carlo simulations.\\ \\
Many classical financial market models are based on the \textit{Efficient Market Hypothesis}(EMH) originally introduced by Fama \cite{malkiel1970efficient, granger1970predictability}.
The EMH faces extensive criticism and controversial discussions are still carried out \cite{malkiel2003efficient}.
One reason for this is the existence of market anomalies, usually named \emph{stylized facts}, which cannot be explained by the EMH.
\textit{Stylized facts} are statistical observations in financial data which can be documented on different time scales and for various stock markets all over the world. 
The \textit{stylized fact} probably best known is the inequality of income and wealth which was first discovered by Pareto in 1897 \cite{pareto1897cours}.
Additional examples are heavy tails in stock return distributions and volatility clustering, originally identified by Mandelbrot in 1963 \cite{mandelbrot1997variation}.
For further discussion of \textit{stylized facts}, we refer to \cite{cont2001empirical, lux2008stochastic}. \\ \\
\textit{Stylized facts} seem to play a major role in the emergence of financial crises \cite{cowan2002heterogenous}
 and the need to investigate the origins of \textit{stylized facts} has been emphasized by several authors \cite{farmer2009economy}. The common goal of ABCEM models is to replicate financial data containing \emph{stylized facts} and thus to discover reasons for their appearance. Hence ABCEM models can help to better understand the emergence of financial crisis. \\ \\
ABCEM models indicate that stylized facts are introduced, for example, by behavioral aspects and psychological misperceptions \cite{cross2005threshold, lux2008stochastic} of agents or learning mechanisms \cite{adam2016stock, ehrentreich2007agent, timmermann1993learning}.
Many ABCEM models are heavily influenced by \emph{behavioral finance} \cite{kahneman2003maps} and do not share many similarities with classical financial market models. 
Thus, the investors within ABCEM models, usually called agents, do not follow the  \textit{homo oeconomicus} \cite{mill1994definition} paradigm of rational utility maiximizers. 
They are rather modeled as \emph{bounded rational agents} in the sense of Simon \cite{simon1955behavioral, simon1957models}.
Furthermore, is the demand of each agent aggregated to the so called excess demand, which is defined as the average of the difference between demand and supply of all agents.
%
In classical economic theory the supply matches demand. This theory is known as \textit{general equilibrium theory} and dates back to John Locke, James Denham-Steuart and Adam Smith.
In the 19th century, the general equilibrium theory has been further developed by Antoine Cournot, Carl Menger and Le\' on Walras.
Probably the most influential model in the general equilibrium theory has been introduced by L\' eon Walras \cite{walras1896elements}.
The model considers an auctioneer, who determines the price in a so called \textit{t\^{a}tonnement} process.
Here, one assumes a \textit{rational market} in the sense that we have perfect information and no transaction costs.
There are further developments of the \textit{general equilibrium theory} due to McKenzie, Arrow and Debreu in the 1950s.
We refer to the book \cite{walker2006walrasian} for a general discussion.
The equilibrium can heuristically be expressed as:
\[
\sum\limits_{i=1}^N v_i^S(S,t)=\sum\limits_{i=1}^N v_i^D(S,t),
\]
where $N\in\N$ denotes the number of market participants and $v^{(\cdot)}(S,t)$ the volume of stocks or assets which each agent demands or respectively offers at a certain price $S$.
The equilibrium price $S_{eq}$ is then given as the price, where the supply matches the demand.
From a mathematical perspective, this results in a fixed point problem, for which the existence of a solution often is not a priori guaranteed and which is usually difficult to solve.\\ \\
The general equilibrium theory is criticized (e.g. \cite{heertje2002recent, ackerman2002still}) due to the restricted nature of the assumption of a \textit{rational} market which seems to be often violated in real world economics.
In addition, there is an ongoing discussion whether market prices represent an equilibrium.
For example, Beja and Hakansson \cite{beja1980dynamic} point out that observed prices are usually not identically to the equilibrium prices.
This happens if the t\^{a}tonnement process is taking too long, such that the computed solution is again in disequilibrium.\\ \\
This has lead to the theory of market prices in disequilibrium \cite{beja1980dynamic, he2011dynamic, day1990bulls} in which the price adjustment speed is finite and the actual market price represents a price in disequilibrium. In fact, most ABCEM models employ the concept of bounded rational agents and consider an irrational market mechanism. \\ \\
In the last decade there have been many contributions of new ABCEM models. We only present a short overview over the most influential ABCEM models:  the Levy-Levy-Solomon model \cite{levy1994microscopic}, the Lux-Marchesi model \cite{lux1999scaling, lux2000volatility}, the Brock-Hommes model \cite{brock1997rational, brock1998heterogeneous}, and the Cont-Bouchaud model \cite{cont2000herd}.
ABCEM models describe a diverse field of applications. Several models focus on the creation of crises (cf. \cite{kim1989investment, kaizoji2000speculative, harras2011grow}) while others try to explore the influence of new regulations of policy makers on the market behavior (cf. \cite{teglio2012impact, da2015combining}).
 We refer interested readers to reviews \cite{lebaron2000agent, bouchaud2002introduction, chakraborti2011econophysics, sornette2014physics, iori2012agent, hommes2006heterogeneous, samanidou2007agent, ehrentreich2007agent, chen2012agent, tesfatsion2006agent, tesfatsion2002agent} for a general introduction to ABCEM models. \\ \\
 %
%
Generally, ABCEM models suffer from the drawback of painting a limited picture of reality.
In fact, an explanation of stylized facts in one model does not necessary hold true in another model.
In addition, critics might argue that all the results are based on computer simulations and cannot be trusted blindly. 
This is a severe issue and earlier studies \cite{egenter1999finite, zschischang2001some, challet2002, kohl1997influence, hellthaler1996influence} have shown that the obtained \textit{stylized facts} in many models are only numerical artifacts.
More precisely, these studies revealed that for example the very influential Lux-Marchesi model and the Levy-Levy-Solomon model exhibit finite size effects \cite{egenter1999finite, zschischang2001some}.
Finite size effects generally describe that different numbers of agents may  lead to qualitative different model outputs.
For that reason it is of paramount importance to simulate ABCEM models with a large number of agents.
Further, Monte Carlo simulations have in general a poor convergence rate and thus a large amount of samples is required to obtain reliable results.
Nevertheless, many ABCEM models are far to complex to study them by analytical methods and therefore computer simulations present the only feasible way. \\ \\
While a huge amount of ABCEM models is presented in literature, to our knowledge a unified model and perspective on ABCEM models is missing. 
Furthermore, there is no objective comparison possible between different models, since the models are implemented in different languages and simulated on different machines.
In addition, we experienced difficulties while reproducing the results published in literature. 
This may have several reasons. First of all ABCEM models are usually non-linear dynamical systems, very sensitive to their parameters. 
Secondly, ABCEM models heavily depend on pseudo random numbers and thus it is impossible to reproduce published results exactly. 
Finally, many publications provide incomplete information regarding the implementation details, e.g. initial values of model quantities.\\ \\
These obstacles motivate us to establish a large scale simulation tool for multi-agent ABCEM models.
Our software tool allows the implementation of many different ABCEM models with a reduced amount of coding. 
This is achieved by providing an object-oriented simulator implemented in C++.
The main building blocks are agents and market mechanisms.
The object-oriented software design enables the user to easily add and test new models by recombining existing agent type or market mechanism implementations using the XML-based configuration mechanism in SABCEMM.\\
We refer to section \ref{features} for further discussion of the possibilities of creating new models with the SABCEMM (Simulator Agent-Based Computational Economic Market Models) tool.
Another advantage of well-implemented C++ code is the computational speed which enables us to run models with several million agents on a Laptop.
This lends SABCEMM particularly well for analysis of statistics of and sensitivity analyses for ABCEM models free of finite size effects.
SABCEMM is built on the novel unified model of ABCEM models derived in section \ref{model}.
We point out that the SABCEMM tool supports numerous pseudo random number generators and enables the user to carry out fair comparisons between different models.
\\ \\
We encourage readers to implement their ABCEM models in our simulator.
Our goal with the SABCEMM tool is to introduce a unified simulator which helps to compare and test ABCEM models.
To aid reproducibility, we publish the code and all examples discussed in this publication under an open source license. \\ \\
We implemented three ABCEM models in our simulator, namely the Levy-Levy-Solomon (LLS) model \cite{levy1994microscopic}, the Cross model \cite{cross2005threshold} and the Harras model \cite{harras2011grow}. We carry out several experiments to analyze the computational efficiency of the SABCEMM simulator. 
Furthermore, we study the impact of different pseudo random number generators on the computational efficiency.
We conducted unit tests and qualitative tests on the model output to verify our implementation. \\ \\  
The outline of the paper is as follows: In section \ref{model} we properly define a unified model from an economic perspective. 
Then we present the SABCEMM software, which is the core part of this paper. 
More precisely, we introduce the software architecture in section \ref{sec-software-architecture} and analyze the SABCEMM software with respect to computational efficiency in section \ref{num-comp-aspects}. This includes scaling behavior and the impact of different pseudo random number generators on the run time of our simulations. We finish this paper with  a short conclusions of this work.

\section{Abstract ABCEM Models}
\label{model}
\begin{figure}[h!]
\begin{center}
\includegraphics[width=0.7\textwidth]{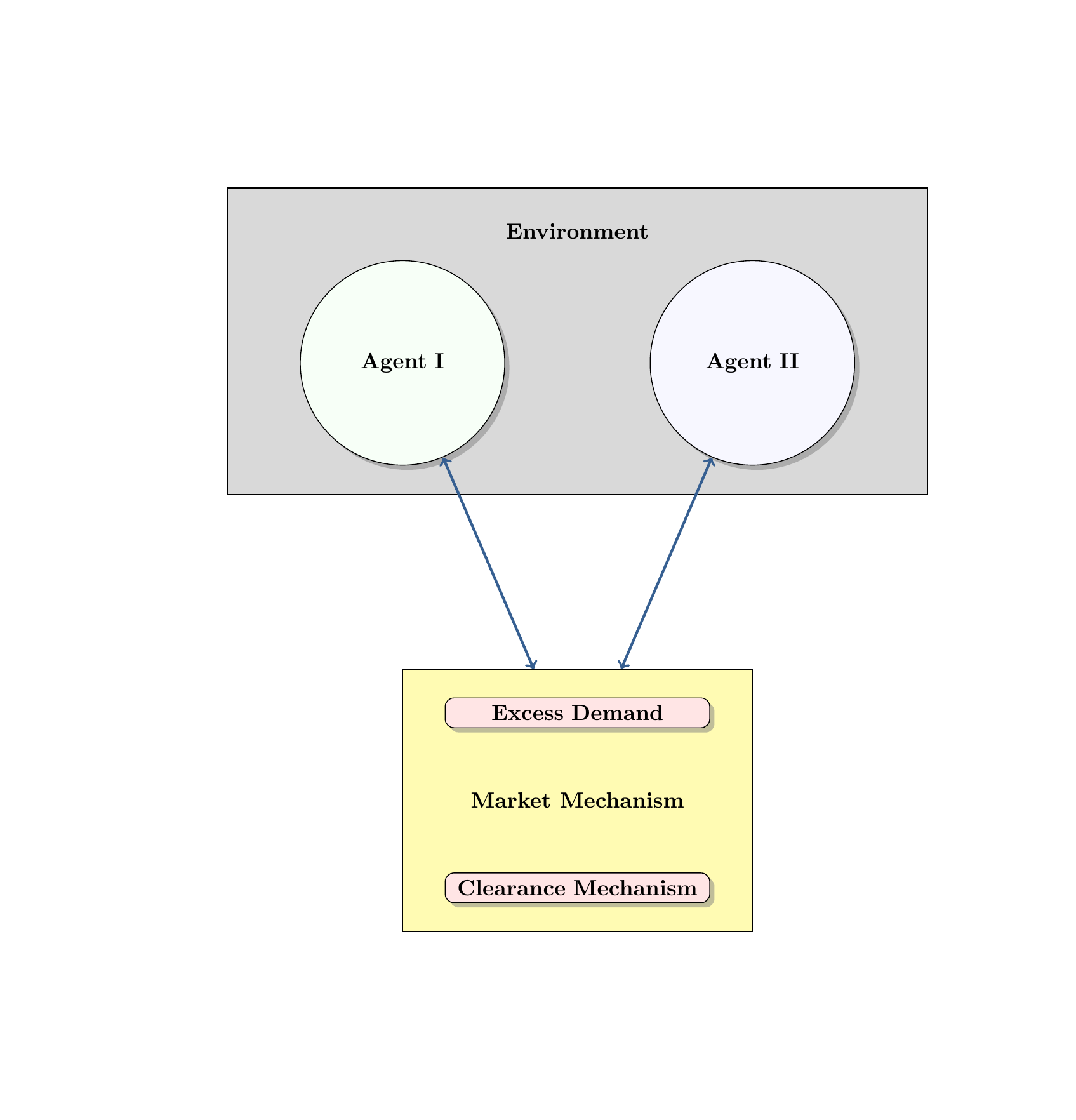}
\caption{Schematic picture of our model. }\label{OurModel}
\end{center}
\end{figure}

In this section, we introduce a meta-model which allows to categorize and to abstract ABCEM models.
All models consist of at least one type of a \emph{financial agent} and a \emph{price adjustment mechanism}. A financial agent is trading at a financial market and interacts through his actions with the price adjustment mechanism.
Here, the agent type describes the strategy an investor follows when acting on a market and interacting with other market participants. 
The other market participant might follow the same or a different strategy, i.e. be of the same or of a different agent type, respectively.
We only consider models with at least two agents,  which even holds true for very stylized models like \cite{cont2000herd} or \cite{brock1997rational}.\\ \\
The obvious second building block is the \emph{market mechanism}, which consists of the \emph{clearance mechanism} and a method of computing the \emph{excess demand}.
Here, the clearance mechanism describes how the price of an asset or a stock in each ABCEM model is adjusted  \cite{hommes2006heterogeneous, lebaron2006agent, tesfatsion2006agent}.
The excess demand denotes the aggregated supply and demand of all agents. 
More precisely the excess demand is defined as the sum of the agents' supplies subtracted from the agents' demands. 
Since the clearance mechanism should equalize supply and demand, it depends on the excess demand of all market participants.
Obviously, the actions of all market participants are coupled through the price, and therefore the excess demand. \\ \\
Finally, we introduce a third aspect called \emph{environment}. 
This is a novel concept which introduces an additional coupling of the agents besides the global market price and provides a general framework for established and novel coupling mechanisms, such as herding behaviour and network topologies of agents implemented in many published ABCEM models \cite{alfarano2005estimation, kirman1993ants, tesfatsion2002agent, schweitzer2009economic}. Though such a component is not a necessary part of an ABCEM model, it is used in many models.
The reason for the prominent role of such an environment is that the additional coupling enables the formation of \textit{stylized facts}.
Our meta-model is outlined in figure \ref{OurModel}.
Before we discuss each building block separately, we present the SABCEMM model.
First, we formally define an agent.
\begin{definition}
An agent is characterized by three aspects:
\begin{enumerate}
	\item a set of individual pieces of information unique to each agent, i.e. any kind of information represented by a set of variables, provided by each agent individually and used in the investment decision process;
	\item a set of public pieces of information, represented by a set of variables, accessible by all agents and used in the investment decision process; and
	\item the investment decision process.
\end{enumerate}
Now, assume a set of $N$ agents, where agent $i$ takes its own set of $J_i$ individual pieces of information, the individual pieces of information of all remaining $N-1$ agents and the set of $J_{ex}$ public pieces of information into account in their investment decision process.
We denote the set of all information used in the investment decision process by $\Omega \subseteq \R^{\sum\limits_{i=1}^N J_i+J_{ex}}$.
The investment decision process is defined by the map:
\begin{align*}
	\left.
		\begin{aligned}
			A_i: \Omega &\to \R^{J_i} \times \R,\\
			\omega &\mapsto A_i( \omega ),
		\end{aligned}
	\right\}
\end{align*}
where the map $A_i$ maps onto a new set of individual pieces of information used in its next investment decision process (representing an update of the set of individual pieces of information) and a price, which needs to coincide for all agents.
\end{definition}
Acknowledging that this definition of agents is rather abstract and general, we provide the following remarks and examples to motivate and clarify this definition.
\begin{remark}
The individual pieces of information are called individual as they are held and updated by the agent providing the individual piece of information only.
Depending on the agent type in the corresponding ABCEM model the individual agent may access individual pieces of information of other agents as well.
Examples of such individual pieces of information may include the opinion on the state of the market, a history of previous investment decisions or the wealth evolution of each individual agent.
\end{remark}
\begin{remark}
The set of public pieces of information may represent publicly available information induced into the market, such as news or the stock price. These information are not individual to each agent. The concept of public pieces of information is introduced with agents as it allows different types of agents to trade at the same market while defining different public pieces of information per type of agents. For example, agents of type A might use public pieces of information representing news from outside the market while agents of type B might use public pieces of information representing the evolution of the stock price over the last $n$ timesteps.
\end{remark}
\begin{remark}\label{NoCoupling}
For agents, where the investment decision process does not depend on individual pieces of information of the other agents, but only on public pieces of information such as excess demand, as used in the Cross model \cite{cross2005threshold}, $\Omega$ reduces to $\Omega \subseteq \R^{J_i+J_{ex}}$ and the investment decision map can be simplified to: 
\begin{align*}
	\left.
		\begin{aligned}
			A_i: \Omega &\to \R^{J_i} \times \R,\\
			\omega &\mapsto A_i( f_{i,1},..,f_{i,J_i}; f_{ex,1},..,f_{ex,J_{ex}} ),
		\end{aligned}
	\right\}
\end{align*}
where $f_{i,j}$ ($j=1,..,J_i$) denote the individual pieces of information of agent $i$ and $f_{ex,j}$ ($j=1,...,J_{ex}$) the public pieces of information, respectively.
Note, that models only including agents of that type are inherently parallelizable.
\end{remark}
\begin{remark}
The effects of the coupling via an environment are implicitly built into the definition of agents, through the possible dependence of the investment decisions of the $i-$th agent on the individual pieces of information of other agents. In the simplified setting of \cref{NoCoupling} there is no coupling present.
\end{remark}
In order to visualize the coupling via an environment, we present the following example.
\begin{example}
Assume a set of $N \in \N$ agents each equipped with a set of four individual pieces of information:
\begin{enumerate}
	\item the wealth of the agent $w_i$;
	\item the number of stock $q_i$ held by the agent;
	\item the investment decision (or opinion) $\sigma_i$ of the agent on the stock;
	\item the average of investment decisions of a subset of agents  $$\bar{\sigma}_i^{m_i,n_i} = \frac{1}{n_i-m_i+1} \sum_{k=m_i}^{n_i} \sigma_k$$ with $m_i,n_i\in {1,..,N} \ \text{and}\ m_i < n_i$.
\end{enumerate}
Here, the average quantity $\bar{\sigma}_i^{m_i,n_i}$, which may depend on the investment decisions of all agents or any subset of agents, introduces the coupling. 
Further, assume that the only external piece of information is given by
\begin{enumerate}
\item the stock price $S$.
\end{enumerate}
Then, the decision process can be written as:
\begin{align*}
	\left.
		\begin{aligned}
			A_i: \Omega &\to \R^{4} \times \R,\\
			\omega &\mapsto A_i\left( S, w_i, q_i, \sigma_i, \bar{\sigma}_i^{m_i,n_i} \right),
		\end{aligned}
	\right\}
\end{align*}
where the map is defined on $\Omega= \R^{4+1}$. 
\end{example}

Based on the definition of agents provided above, we now introduce the notion of agent types allowing for an easier discussion of ABCEM models.
\begin{definition}
We define when two agents $A_i$ and $A_j$ with $i \neq j$ are of the same agent type. The maps $A_i$ and $A_j$ are identical, if it is possible to obtain one of the other by a simple permutation of the input vector $\omega$.
\end{definition}
Next, we define the market mechanism consisting of the \emph{excess demand} and the \emph{clearance mechanism}.
\begin{definition}
The aggregated excess demand $ED$ of $N$ agents is defined as the average of the agents' microscopic excess demands $ed_i\in \R$:
$$
ED(S):= \frac{1}{N} \sum\limits_{i=1}^N ed_i(S). 
$$
The agent's excess demand $ed_i$ depends on the stock price $S$, thus $ed_i=0$ corresponds to no orders of the i-th agent.
A positive value $ed_i>0$ represents a  buy order, whereas a negative $ed_i<0$ reflects a sell order. 
Our meta-model does not prescribe the explicit form of $ed_i$ which is specific to each ABCEM model.
For a detailed discussion of aggregated excess demand we refer to \cite{mantel1974characterization, sonnenschein1972market}
\end{definition}

\begin{example}\label{exampleED}
In \cite{harras2011grow} a reasonable choice for $ed_i(S)$ is given by:
$$
ed_i(S)= \sigma_i(S) \frac{\gamma_i(S)\ w_i(S)}{S},
$$
with investment position $\sigma_i \in\{-1,0,1\}$, investment fraction $\gamma_i\in [0,1]$, wealth $w_i\geq 0$ and market price $S$.  
\end{example}

\begin{definition}
    \label{defitiontionMarkets}
The SABCEMM simulator provides two possible clearance mechanism.
\begin{itemize}
\item[i)] Rational market: 
$$
ED(S)\stackrel{!}{=}0.
$$
\item[ii)] Irrational market: 
$$
S_{k+1}= M( S_k,  ED, \eta).
$$
\end{itemize}
Here, the subscript $k\in \N$ denotes a time step in a time discretization of a time period $[0,T],\ T>0$.
By $\overset{!}{=}$ in the definition of the rational market, we denote that $S$ needs to be chosen such that the equation is satisfied, i.e. $S$ is implicitly defined by the equation.
\end{definition}

Note that the rational market is a root-finding problem, whereas the irrational market is a numerical approximation of a differential equation. 
The general form of the irrational market (defined by the function $M$) does not only include explicit discretization schemes but may also include exponential integrators to approximate the equation.

\begin{example}
When using the agent's excess demand as defined in example \ref{exampleED}, we obtain as an example of a rational market the following equation:
$$
0\stackrel{!}{=}\frac{1}{N}\sum\limits_{i=1}^N\sigma_i(S) \frac{\gamma_i(S)\ w_i(S)}{S}.
$$
An example of an irrational market reads
$$
S_{k+1}= S_k+  \frac{1}{N\ \lambda} \sum\limits_{i=1}^N\sigma_i(S) \frac{\gamma_i(S)\ w_i(S)}{S}.
$$
Here, $\lambda>0$ denotes the market depth \cite{kempf1999market} and the previous price setting rule has been used in the model \cite{harras2011grow}.
\end{example}
We refer to the appendix \ref{software-classification} for an example how to translate our abstract meta-model into software instructions. 
In the remainder of this section, we discuss each building block separately in order to provide insight into the theoretical background of each modeling aspect.
In addition, we provide some examples of popular agent types and market mechanisms in ABCEM models. 

\subsection{Market Mechanisms}
\label{different-market-mechanism}
In this section we motivate the market mechanism of our meta-model. We aim to illustrate the connection between a rational and an irrational market. 
In order to do so, we present the disequilibrium model of Beja and Goldman \cite{beja1980dynamic}. As central concept we need to understand the meaning of \textit{aggregated excess demand} $ED$ (see for example \cite{mantel1974characterization, debreu1974excess, sonnenschein1972market}).
In essence the aggregated excess demand denotes the sum of the demand subtracted by the sum of the supply.    
A positive aggregated excess demand represents non cleared buy orders and a negative aggregated excess demand non cleared sell orders.\\ \\
The disequilibrium model by Beja and Goldman \cite{beja1980dynamic} reads: 
\begin{align}
\frac{d}{dt} P=H(ED(P)),\quad P:=\log(S),\label{BG}
\end{align}
where $S$ denotes the price of the stock, bond, future or option and $ED$ the \textit{aggregated excess demand} and $\log(\cdot)$ denotes the Napierian logarithm.
Furthermore, the function $H$ is assumed to be a monotone increasing function, which vanishes at zero.
The function $H$ might be nonlinear which is supported by several studies \cite{campbell1997econometrics, kempf1999market, cont2000herd}.
Beja and Goldman approximate model \eqref{BG} by a first order linearization of $H$ (Taylor expansion of $H$ with $\dot{H}(0)=\frac{1}{\lambda}$):
\begin{align}
\frac{d}{dt} P=\frac{1}{\lambda} ED(P),\label{BGsimple}
\end{align}
where the constant $\lambda>0$ is interpreted as the market depth \cite{kempf1999market}.
Mathematically, such a linearization of the function $H$ is a good approximation for small values of $ED$.
In fact, studies of Farmer et al. \cite{cont2000herd, farmer1997} indicate a linear trading impact for small price changes.
Beja and Goldman add white noise to their model \eqref{BGsimple} to cover random errors or external news.
Hence, the clearance mechanism is given by:
\begin{align}
\frac{d}{dt} P=\frac{1}{\lambda} ED(P)+\eta,\quad \eta\sim\mathcal{N}(0,1). \label{BGSDE}
\end{align}
Thus the ordinary differential equation \eqref{BG} has become the stochastic differential equation \eqref{BGSDE}.
The stochastic differential equation is properly defined as an integral equation and can be interpreted in the It\^{ o} or Stratonovich sense (cmp. \cref{technicalDetails}). \\ \\
Mathematically, the process of transforming the algebraic demand supply equation
\begin{align}
0=ED(P).\label{algprice}
\end{align}
 into the differential equation \eqref{BGsimple} can be interpreted as relaxation, where the rate of relaxation is given by the market depth $\lambda$.
The excess demand is usually measured in wealth or number of stocks.
Thus, the right hand side of \eqref{BGsimple} is a rate due to the multiplication by the market depth.
Hence, the most general model for a stochastic differential equation is given by
 \begin{align}
 dS = F(S, ED)\ dt+ G(S,ED)\ dW, \label{GenMod}
 \end{align}
 with Wiener process $W$(cmp. \cref{technicalDetails}) and arbitrary functions $F$ and $G$.
 Notice, that \eqref{BGSDE} is a special case of the model \eqref{GenMod}.
 We use the usual notation for  It\^{o} stochastic differential equations.
 Many market mechanism of ABCEM models are special cases of model \eqref{GenMod}, for example the models presented in \cite{day1990bulls, alfarano2008time, lux1995herd, chiarella2002speculative, chiarella2005dynamic, chiarella2006asset, chiarella2007heterogeneous, challet2001stylized, zhou2007self, andersen2003game, harras2011grow, sornette2006importance, kaizoji2002dynamics, palmer1994artificial, bouchaud1998langevin, cont2000herd, cross2005threshold, cross2007stylized, cross2006mean, dieci2006market, farmer2002price, lux1999scaling, lux2000volatility, de2005heterogeneity}.  
 
 \paragraph{Discretization}
 All ABCEM models can be regarded as a time discrete versions of time continuous models. 
One needs to discretize time continuous models in order to be able to implement and simulate the corresponding numerical approximation on a computer (cmp. \cref{technicalDetails}).

In the ABCEM literature one usually finds explicit Euler discretizations. 
Often, the numerical approximation (cmp. \cref{technicalDetails}) is rescaled and fixed such that the time step is set to one.
Hence, in ABCEM literature, we are rather faced with difference equations of the following type
   \begin{align}\label{GenModApp}
 S_{k+1} = S_k+  \bar{F}(S_k, ED_k)+ \bar{G}(S_k,ED_k)\ \eta,
 \end{align}
than with differential equations. The model \eqref{GenModApp} is a discretized version of the model \eqref{GenMod}. The functions $\bar{F}, \bar{G}$ represent discretized versions of functions $F, G$. Here $k\in\N$ is an index of the discretized time steps ($S_k=S(t+k\ \Delta t)$ for a fixed initial time $t$ and time step $\Delta t>0$).

\subsection{Agent Design}
In this section, we discuss the design of agents for ABCEM models. The agents are often designed as \textit{bounded rational agents} in the sense of Simon \cite{simon1955behavioral, simon1957models}. This means that the investors rather build their investment decisions on heuristics (behavioral rules) than a perfect utility maximization. Mathematically, they do not solve an optimization problem, but derive a satisfactory solution near the optimum by their trading rules. Such suboptimal trading strategies are astonishingly good approximations of the real investment process \cite{ehrentreich2007agent, tiwana2007bounded}. Furthermore, the heuristic trading rules often incorporate psychological aspects in the investment decision. For an introduction to the discipline \emph{behavioral finance} we refer the interested reader to an article by Kahnemann \cite{kahneman2003maps}. We want to point out that agents in ABCEM models may also be build on other concepts, for example, the so called zero-intelligence trader \cite{farmer2005predictive, gode1993allocative}. The SABCEMM simulator does not make any limitations regarding the modeling of an agent. \\ \\
As an illustrative example of heuristic trading strategies we present two frequently used investor types \cite{lux1998socio, hommes2006heterogeneous}: \emph{chartists} (technical trader) and \emph{fundamentalists}.
A fundamentalist investor believes that there exists a fair price for an asset or a stock and that the market price will converge to this value. For a given fundamental value $S^f$ and a monotonically increasing function $D_i$ one may define
\begin{align}
ed^F_i(S):=D_i(S_i^f-S). 
\end{align}
In contrast to the fundamentalist, the \textit{chartists} forecast the future price by extrapolating past values. In the simplest setting the chartist may only consider the last stock prices.
\begin{align}
ed^C_i(S):=D_i(S_k-S_{k-1}).
\end{align}
The previous definition are generalizations of the choices made in  \cite{lux1998socio, hommes2006heterogeneous}.
Examples of ABCEM models, which consider chartists or fundamentalists are \cite{levy2000microscopic, chiarella2006asset, brock1997rational, brock1998heterogeneous, chiarella2006asset, franke2012structural}.

\subsection{Environment}
In this section, we introduce the third aspect of our meta-model: the environment.
An alternative name for the environment is coupling and it represents the crucial ingredient of many ABCEM models. \\ \\
A frequently used environment is the \textit{herding mechanism}.
Kirman \cite{kirman1993ants} was possibly the first who used herding.
Herding makes investors flock together and creates high correlations among the financial agents.
This leads to rapid up or down movement in the market price and non-Gaussian price behavior.
Several models have implemented the herding mechanism e.g. \cite{alfarano2005estimation, kirman2001microeconomic, kirman1993ants, cross2005threshold, franke2012structural}.
\begin{example} As an example we present the herding mechanism of the Cross model \cite{cross2005threshold}. Each agent is described by a herding pressure $c_i>0$ and the evolution reads
\[ \begin{cases} c_i(t+\Delta t)= c_i(t)+ \Delta t |ED(t)|,& \text{if}\ \sigma_i(t)\ ED(t)<0\\
			c_i(t+\Delta t)=c_i(t),& \text{otherwise}.
			 \end{cases}
\]
Thus, the herding pressure is increased if the investment decision of the agent $\sigma_i\in\{-1,1\}$ has the opposite sign as the aggregated excess demand $ED$. 
This situation corresponds to the fact that the agents' position is in the minority. 
The agent switches position if the herding pressure $c_i$ has reached a threshold $\alpha_i>0$. After a switch the herding pressure gets reset to zero. 
This herding mechanism leads to additional coupling beneath the agents, introduced by the coupling with the aggregated excess demand. 
For a further presentation of the model we refer to the appendix \ref{appendixModel}.
\end{example}

The second frequently used coupling mechanism in ABCEM models is a switching mechanism between different agent groups. 
Switching allows agents to change investment strategy resulting in a varying weight of implemented investment strategies.
Thus the price behavior is often mainly influenced by one investment strategy. 
The switch or more precisely the switching rate is often triggered by a \textit{fitness measure}. The \textit{fitness measure} of an investor, or the investor group, is usually a comparison of past or actual profits of the different investment strategies. Thus, such a switching mechanism again creates additional correlation among agents.
Prominent examples are \cite{brock1997rational, brock1998heterogeneous, franke2012structural, lux1999scaling}. \\ \\
Nevertheless we have to point out that there are other environments in ABCEM models which seem to create stylizes facts. Examples are agent interactions on lattice topologies or couplings through global information streams \cite{zhou2007self, harras2011grow}.

\section{The SABCEMM Software}
\label{sec-software-architecture}
In this section, we provide a bird's eye view on our simulation software {SABCEMM} (Simulate Agent Based Computational Economic Market Models).
It allows a straight forward implementation of the general ABCEM model  introduced in \cref{model}.
First, in \cref{numerical-core}, the building blocks of the numerical core of the {SABCEMM} simulation software are laid out.
Then, we note how to build new models out of the existing codebase in order to create new models with almost no additional coding in \cref{features}.
As a next step we discuss the importance of software testing and the implementation in this software project \cref{testing}.
Then, in \cref{num-comp-aspects}, we present results for scalability and efficiency of our code showing that the code is suited for simulation of models including more than $3$ millions of agents.
Finally, we discuss the impact of different pseudo random number generators on the simulation time \cref{RandomNumbers}.
All simulation results can be downloaded at \cite{DataS}.
\\ \\
An object-oriented design of the {SABCEMM} simulation software allows the implementation of Dijkstra's \cite{dijkstra1982role} principle of \emph{separation of concerns}.
More precisely, we follow a class based object-oriented programming approach as described in \cite{rumbaugh1991object}.
In our specific case the object-orientation enables the user to test and implement new econophysical models based on already existing building blocks with minimal additional coding in the {SABCEMM} simulator.\\
The code is documented in the  \texttt{Reference Manual}\footnote{\url{https://sabcemm.github.io/SABCEMM/}} and we provide a  \texttt{User Guide}\footnote{\url{https://github.com/SABCEMM/SABCEMM/wiki/User-Guide}} to facilitate the use of our software tool.
The full codebase can be downloaded on GitHub \url{https://github.com/SABCEMM/SABCEMM} and may serve as an example for further code development by the reader.
In addition, we provide \footnote{ Cross herding mechanism: \url{https://github.com/SABCEMM/SABCEMM/blob/v0.1-alpha/src/Agent/AgentCross.cpp\#L138-L140}} \footnote{Harras ED: \url{https://github.com/SABCEMM/SABCEMM/blob/v0.1-alpha/src/ExcessDemandCalculator/ExcessDemandCalculatorHarras.cpp\#L71-L87}} links to code excerpts in order to present the translation of agent dynamics into C++ code. 
In \cref{software-classification} we classify our software by the categories proposed in \cite{nikolai2009}.

\subsection{Numerical Core}
\label{numerical-core}

\begin{figure}[h!]
    \begin{center}
        \includegraphics[width=0.7\textwidth]{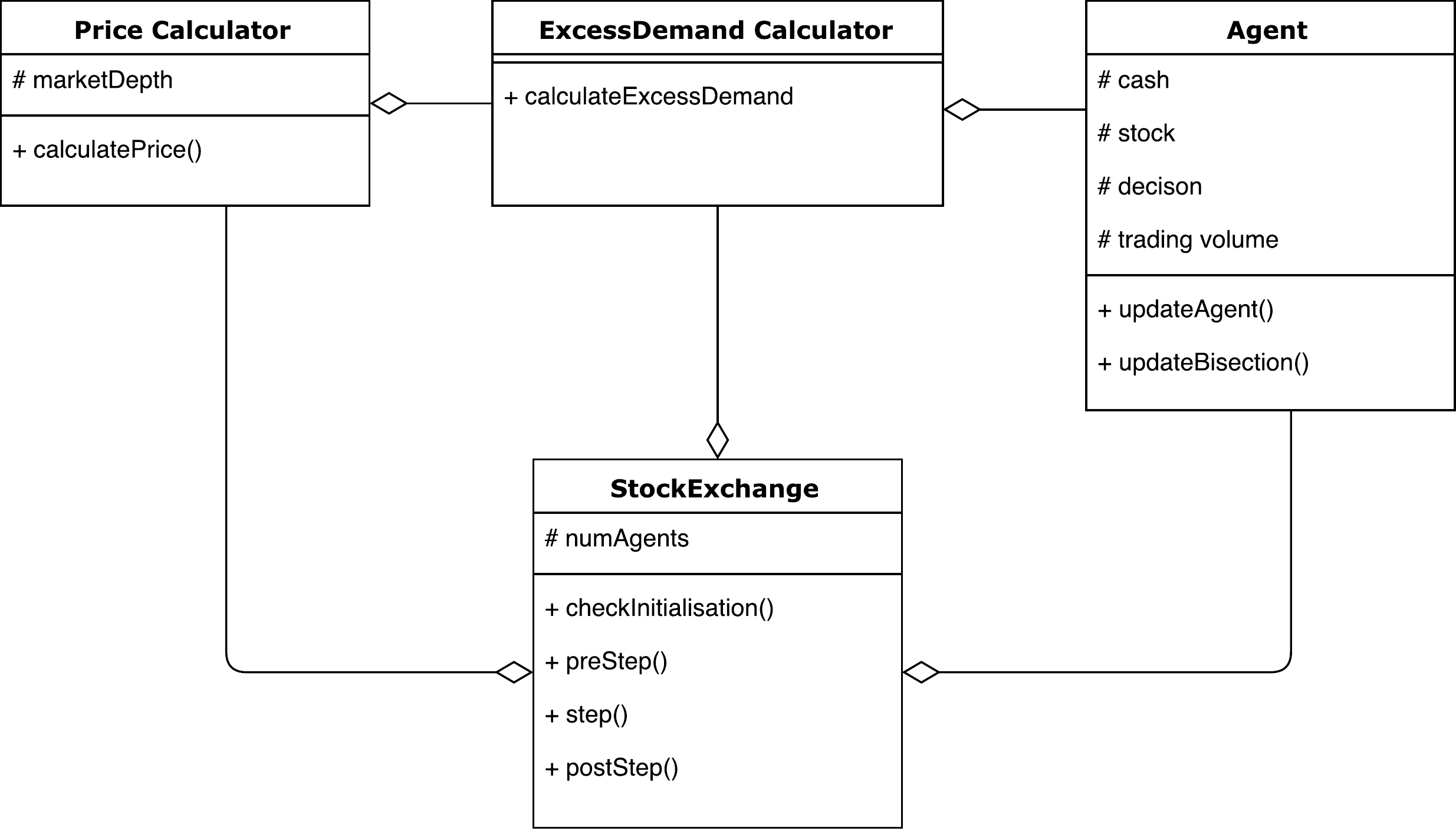}
        \caption{Class diagram of the numerical core. }
        \label{classDiagrammNummericalCore}
    \end{center}
\end{figure}

In this section, we present the design of the numerical core of the {SABCEMM} simulation software, i.e. the design of those parts of the software used in the simulation loop itself.
The three main building blocks of the numerical core and the main workhorses of the simulation are abstracted into abstract classes: \texttt{Agent}, \texttt{ExcessDemandCalculator} and \texttt{PriceCalculator}.
The model to be simulated is then implemented using specialized subclasses of these building blocks vastly reducing the cost for implementation of new econophysical models.
The interaction of the building blocks is orchestrated by the class \texttt{StockExchange}.
The building blocks of the numerical core are visualized by a class diagram in figure \ref{classDiagrammNummericalCore}.

\begin{figure}[h!]
	\begin{center}
		\includegraphics[width=0.5\textwidth]{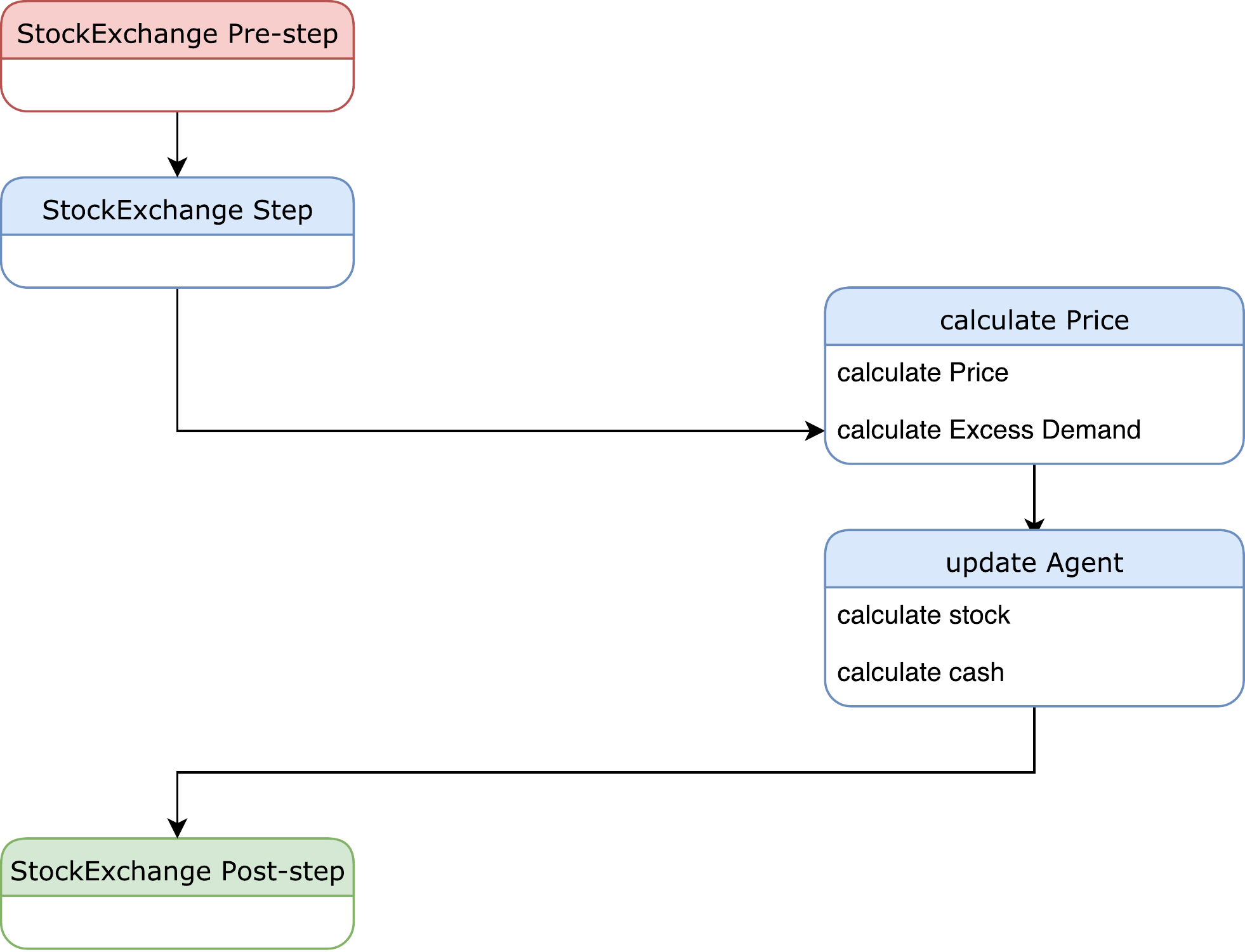}
		\caption{Flow chart of the numerical core. }
		\label{flowChartNumericalCore}
	\end{center}
\end{figure}

\paragraph{The Abstract Class \texttt{Agent}}
The abstract class \texttt{Agent} defines the general interface, i.e. those general characteristics of all agent types required for simulation.
In order to comply to \cite{harras2011grow,cross2005threshold,levy1994microscopic, levy1995microscopic}, every agent type needs the following member variables:
\begin{itemize}
    \item \texttt{Stock} is the amount of stocks an agent holds at a distinct time. In more sophisticated models like \cite{harras2011grow} agents can buy/sell varying amounts of stock.	
    \item \texttt{Cash} is the amount of money the agent has available. In simple models this is called wealth since there is no distinction between stocks owned and disposable cash. 
    \item The \texttt{Decision} can generally take the three discrete values $\{-1,0,1\}$ which stand for sell/no/buy order. 
    \item \texttt{Trading Volume} is the amount of stocks the agent wants to buy or sell.
\end{itemize}
Noted that not all characteristics are relevant for all models, i.e. some might be set  to one/zero to avoid repetition of code.
In addition, every agent type needs to implement the following methods:
\begin{itemize}
    \item \texttt{updateAgent()} does all necessary computations to migrate the agent from one time step to the next one, e.g. revise his \texttt{Decision} and update  \texttt{Cash}/ \texttt{Stock}.
    \item \texttt{updateBisection()} is needed in a rational market only. While searching the next price some of the agent's quantities have to be adapted while others have to remain unchanged.
\end{itemize}
Note that for models with agents relying on an environment (see \cref{model}) the environment needs to be integrated into the specialization of the \texttt{Agent} class.
\begin{example}
    The agents as defined by \cite{harras2011grow} are grouped on a virtual square lattice with periodic boundary conditions, such that each agent has four neighbors (for details see \cref{appendix-harras-model}). Consequently the corresponding specialization of the \texttt{Agent} class can manage a neighborhood.
\end{example}
\paragraph{The Abstract Class \texttt{PriceCalculator}}
defines the general interface of all implementations of the computation of the new price of a stock.
The method \texttt{calculatePrice} determines the new stock price at each time step.
Each of the price mechanisms presented in \cref{model} is implemented in a subclass of the abstract class \texttt{PriceCalculator}.
\paragraph{The Abstract Class \texttt{ExcessDemandCalculator}}
describes the interface to every class implementing a method for calculating the excess demand in a model (compare \cref{model}).
The Excess Demand represents the coupling element between the agent and the price. 
The method \texttt{calculateExcessDemand} iterates over all agents and collects their microscopic excess demand to calculate the global excess demand.
Note that the \texttt{PriceCalculator} relies on the excess demand to find the new stock price.
\paragraph{The Class \texttt{StockExchange}}
represents the interaction between the agents, the price calculator and the excess demand calculator. 
Its interface includes the following member variable and methods:
\begin{itemize}
    \item Member variable \texttt{Agents} represents a list containing all agents trading at a stock exchange.
    \item Method \texttt{preStep()} is called before a step is carried out. It allows implementation of housekeeping tasks, such as collecting data for tracking before a time step is carried out.
    \item Method \texttt{postStep()} is called after the time step is carried out. It allows implementation of housekeeping tasks, such as collecting data for tracking after a time step is carried out.
    \item In method \texttt{step()}, the price calcuator is called to determine a new price. This invokes the excess demand calculator first to determine the excess demand. With the new price all agents are then updated.
\end{itemize}
Figure \ref{flowChartNumericalCore} shows a flow chart of how the different classes work together.
Up to now all investigated models can be rewritten to rely on a single \texttt{StockExchange} class and are then consistent with the flow in \cref{flowChartNumericalCore}.
%
\subsection{Building New Models}
\label{features}
The goal of the {SABCEMM} simulator is to allow for simple implementation of different ABCEM models while permitting fast simulations with larger numbers of agents and providing easy access to simulation results for evaluation of the implemented models.

\paragraph{Building Blocks}
All implementations of the abstract classes \texttt{PriceCalculator},  \texttt{Agent} and \texttt{ExcessDemandCalculator} are building blocks to (new) models. 
So far we implemented all necessary blocks for the Harras \cite{harras2011grow}, LLS \cite{levy1994microscopic, levy1995microscopic} and Cross \cite{cross2005threshold} model which behave as defined in \cref{appendixModel}.
In principle all blocks can be recombined to the user wishes.
This is due to the object oriented architecture and a main advantage of the SABCEMM software.
The interaction between the blocks is defined by abstract interfaces.
If a model can be reformulated as an abstract ABCEM model as defined in \cref{model}, then it can be implemented with existing or new building blocks.
Usually a block then requires a set of parameters which is provided in the input file.\\
While the building blocks provide great flexibility, it is the user's task to determine whether the chosen combination of blocks and parameters form a valid ABCEM model and produce scientifically relevant results.\\
Information on all available building blocks can be found in the SABCEMM documentation\footnote{\url{https://github.com/SABCEMM/SABCEMM/wiki/Create-an-Input-File}}.

\paragraph{Configuration via Input Files}
In order to evaluate the combination of different blocks or examine different parameter settings, a large number of simulations is required.
Additionally a single simulation has to be repeated multiple times to analyze the influence of randomness on the simulation results. 
The combination of building blocks defining a model for a simulation and their respective parameters are specified in \texttt{XML} formatted input files.
The configuration files are well structured, human editable, and are well suited for version control systems allowing a reproducible workflow.
Parameter studies can easily be carried out by using scripting languages for assembly of the required input files.
\begin{example}
    To replace the price mechanism of the Cross model \cite{cross2005threshold} by a rational market (cmp. \cref{defitiontionMarkets}) we change the \texttt{priceCalculatorSettings} section in the XML input file from
\begin{lstlisting}[frame=single]
...
<priceCalculatorSettings>
    <priceCalculatorClass>
        PriceCalculatorCross
    </priceCalculatorClass>
    <theta>2</theta>
    <marketDepth>0.2</marketDepth>
</priceCalculatorSettings
...
\end{lstlisting}
to the following:
\begin{lstlisting}[frame=single]
...
<priceCalculatorSettings>
    <priceCalculatorClass>
        PriceCalculatorBisection
    </priceCalculatorClass>
    <epsilon>0.1</epsilon>
    <maxIterations>10000</maxIterations>
    <lowerBound>0.01</lowerBound>
    <upperBound>200</upperBound>
    <theta>2</theta>
    <marketDepth>0.2</marketDepth>
</priceCalculatorSettings>
...
\end{lstlisting}
Here we configure a bisection method to approximate the price for the next time step.
The full input file can be found on GitHub\footnote{\url{https://github.com/SABCEMM/SABCEMM/blob/master/input/examples/Cross.xml}} and the parameters are explained in the documentation on building blocks.
\end{example}

\begin{figure}[h!]
    \begin{center}
        \includegraphics[width=0.5\textwidth]{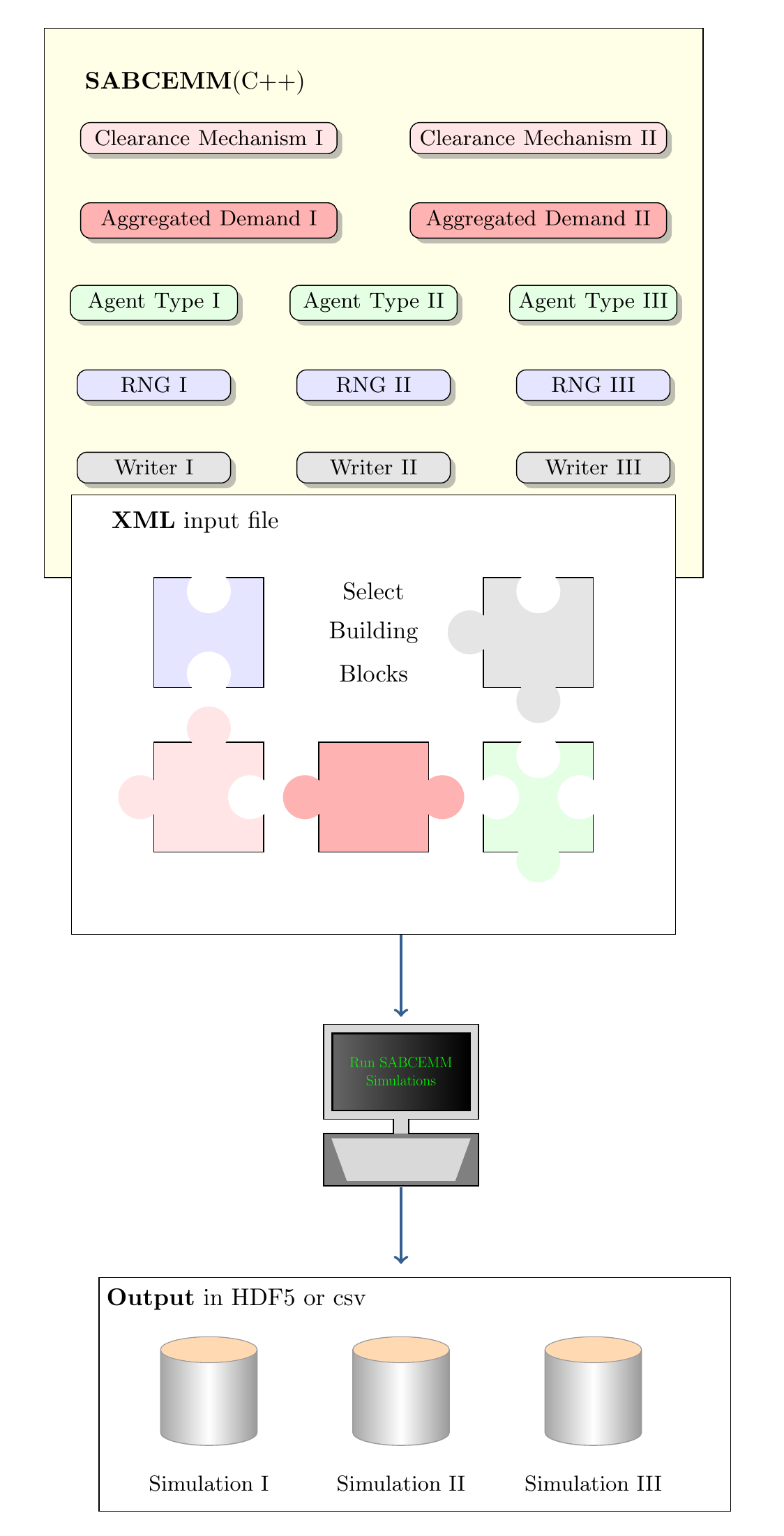}
        \caption{Schematic summary of the structure of the SABCEMM simulator.}
        \label{SABCEMMworkflow}
    \end{center}
\end{figure}

\paragraph{Output} To evaluate possibly thousands of simulations, it is important to write the output in a way such that it is later well accessible for analysis.
SABCEMM offers two formats for output files.
In the basic version, everything is stored to \texttt{csv} files in a dedicated folder.
This method does not rely on third party software and is available on all computers.
A more sophisticated possibility is to store the entire results of a simulation to an \texttt{HDF5} file.
An \texttt{HDF5} file offers an internal structure for data similar to a file system and is a self-describing format.
This allows proper readability of the results stored in an \texttt{HDF5} file.
In addition, the \texttt{HDF5} format allows to also store the input \texttt{XML} used for a simulation within the \texttt{HDF5} output file.
Hence, the output contains all the information necessary to analyze an ABCEM model which aids in ensuring correct documentation of simulation results.
The \texttt{HDF5} file format can be read using the \texttt{HDF5} libraries and utilities\footnote{\url{https://support.hdfgroup.org/products/hdf5_tools/index.html}}, Python via \texttt{h5py}\footnote{\url{https://www.h5py.org}}, MATLAB and Excel via the add-in \texttt{PyHexad}\footnote{\url{https://github.com/HDFGroup/PyHexad}}.
We do not provide any routines for visualization or analysis of the results of simulations as proper post-processing of simulation data is individual to the research carried out.\\ \\
\paragraph{Summary}
Figure \ref{SABCEMMworkflow} illustrates the use of the SABCEMM simulator. SABCEMM consist of building blocks implemented in C++, such as different clearance mechanisms or different agent types.
In Figure \ref{SABCEMMworkflow}, the different types of building blocks are color-coded.
In order to build an ABCEM model from these building blocks for a simulation, one combines the respective building blocks, similar to pieces of a puzzle, using the XML input file.
Additional simulation parameters, such as the number of time steps and the number of simulation runs, e.g. for statistics, are also defined within the XML input file.
Note that our simulator requires the selection of a single market mechanism, aggregated demand, data writer and random number generator but allows for the selection of multiple agent types.
The simulation is carried out by providing the input file to the SABCEMM simulator.
The output can be stored in \texttt{csv} or HDF5 file format. In the case of several runs, e.g. in the case of different parameters or repetitions of the same setting, the simulation output is stored in separate files. 

\subsection{Testing}
\label{testing}
With ever increasing complexity of software systems, the importance of software testing has become more and more evident. This has led to great advances in testing theory \cite{myers2011art}.
Generally, in large scale and complex problems bugs can be mistaken for features of the simulation and vice versa.
Therefore, the reliability of simulation results depends on the reliability of the software used to obtain the results.
This reliability cannot be completely achieved using thorough testing but approximated to a high degree. 
 In SABCEMM, test are implemented using the \texttt{GoogleTest} library \cite{googletest}.
Usually, numerical simulations are tested against their respective analytical solutions.
However, finding analytical solutions for ABCEM models is only possible for extremely simplified settings.
Note that correctness for simplified settings does not guarantee correctness for general cases.
Additionally, the models we consider heavily rely on pseudo random numbers.

Therefore, we use a different approach to testing. 
\begin{itemize}
    \item 
    \textbf{Unit Tests} execute  the code of a single function.
    For these tests all variables are initialized and the end-result of the function is compared to the expected result.
    \item 
    \textbf{Integration and Acceptance Tests} cover multiple functions.
    Thus they are especially useful to test a full time step in our simulator.
    For each model, we calculate one time step by hand.
    This is done once only during the development of the test.
    Afterwards the model output is compared automatically to these values.
    Whenever the model draws a pseudo random number, we usually use the same constant for each agent or use a deterministic number generator (we could draw numbers linearly from a range or similar).
    We use a minimum number of agents to keep time and effort low. 
    %
    %
    \item \textbf{Qualitative Comparison} with earlier research papers.
    As a weaker criterion, we also compare our results to plots given in papers of the corresponding models.
    As typical models rely heavily on random numbers, results from literature can never be reproduced exactly. 
    This is why only a qualitative comparison is feasible.
\end{itemize}
While testing cannot prove the absence of mistakes in the implementation, the combination of all three testing approaches has lead to trustworthy results. We therefore believe that testing is of major importance in any software project.

\subsection{Computational Aspects} 
\label{num-comp-aspects}
\begin{figure}[h!]
    \begin{center}
        \includegraphics[width=0.7\textwidth]{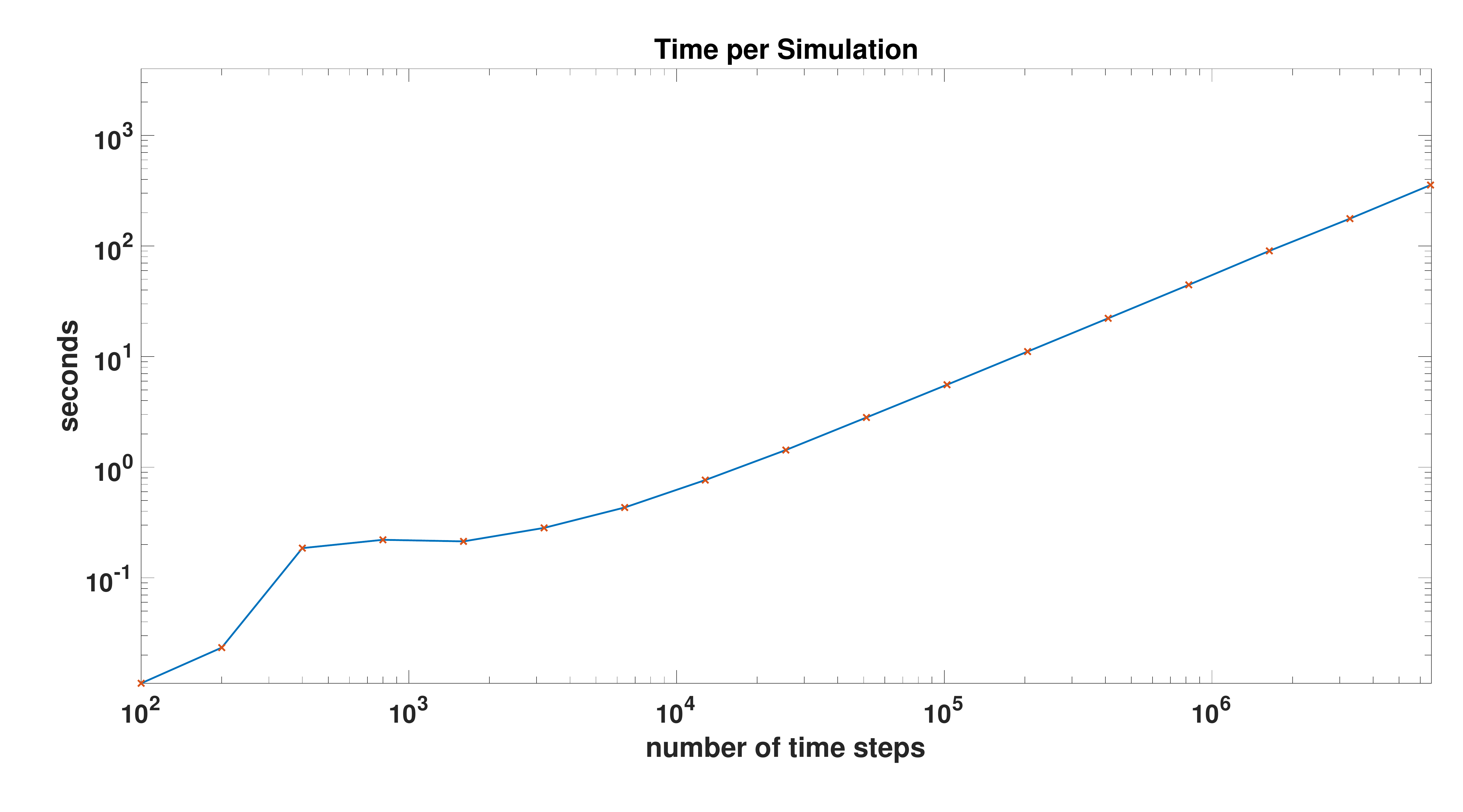}
        \caption{Scaling of the Cross Model with respect to the number of time steps. Parameters as in \cref{cross-basic-parameter}. The time steps are varied according to the plot.}
        \label{scaling-timestepsWMatlabloglog}
    \end{center}
\end{figure}

\begin{figure}[h!]
    \begin{center}
        \includegraphics[width=0.7\textwidth]{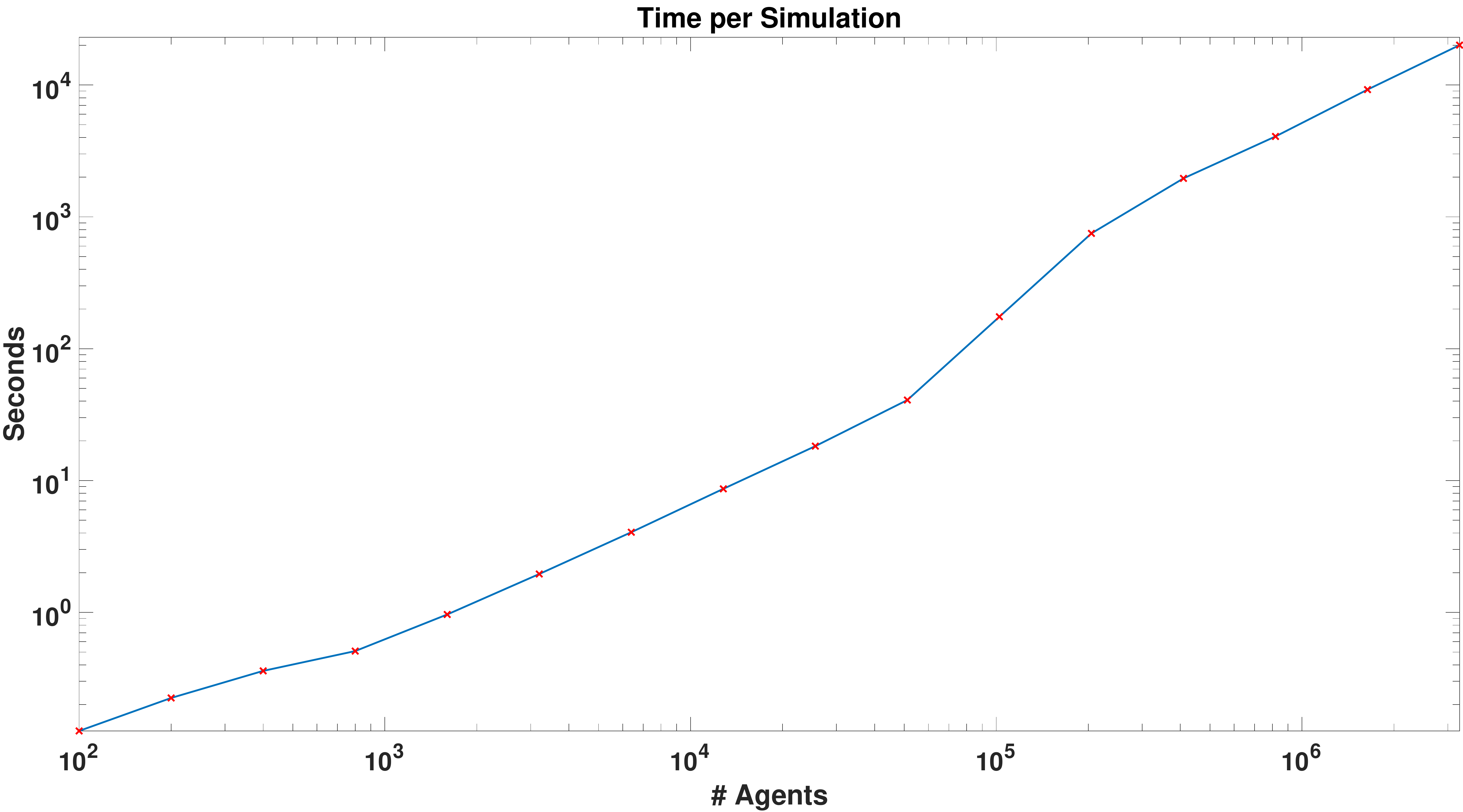}
        \caption{Scaling of the Cross model with respect to the number of agents. Parameters as in \cref{cross-basic-parameter}. The parameter $N$ is varied according to the plot.}
        \label{scaling-agentsloglog}
    \end{center}
\end{figure}

\begin{figure}[h!]
    \begin{center}
        \includegraphics[width=0.7\textwidth]{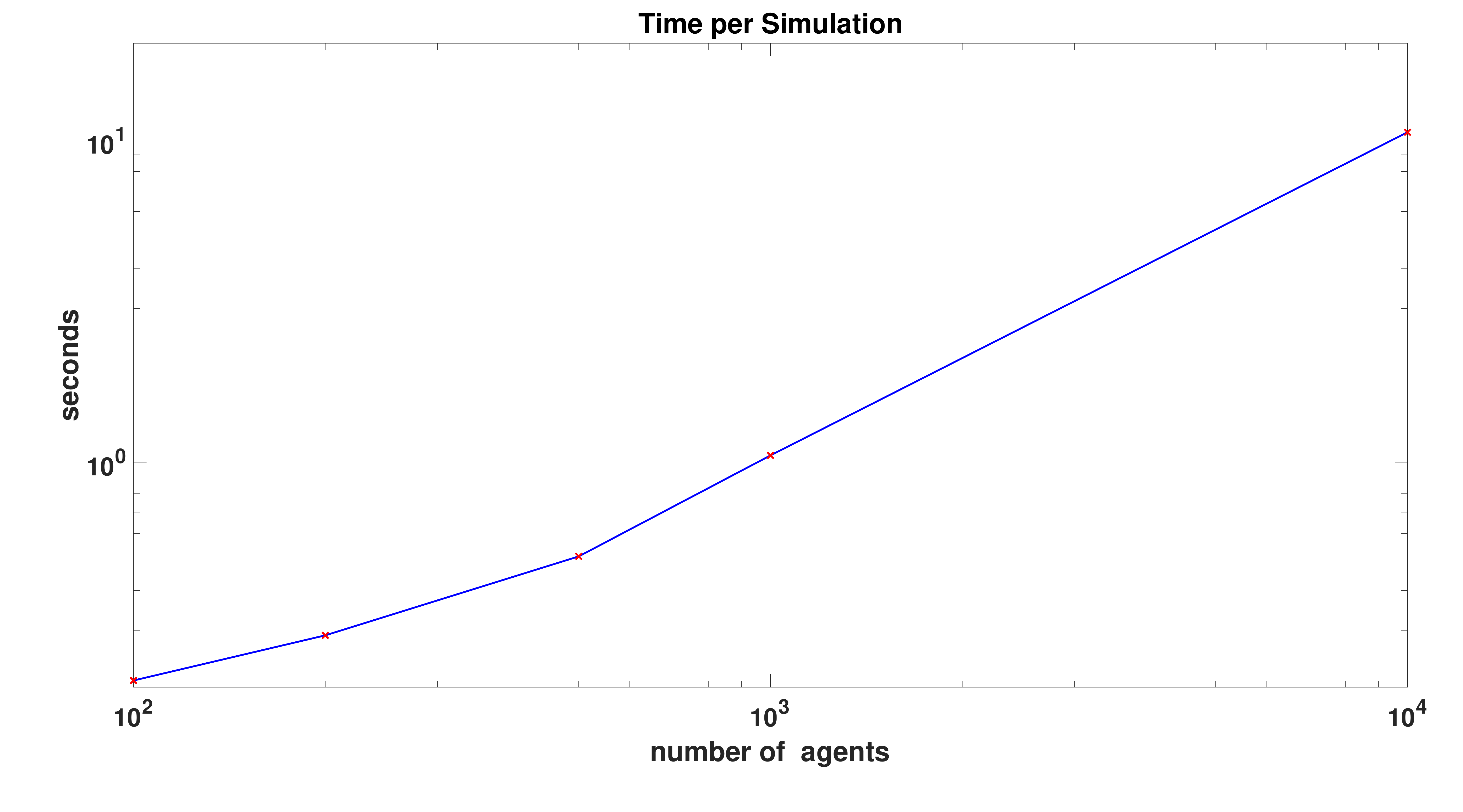}
        \caption{Scaling of the LLS model with respect to the number of agents. Parameters as in \cref{LLS-basic} with $\sigma_{\gamma}=0.2$. The parameter $N$ is varied according to the plot.}
        \label{scaling-agentsLLS}
    \end{center}
\end{figure}

We now analyze how the runtime of simulations scales with the number of agents and number of time steps. 
From figures \ref{scaling-timestepsWMatlabloglog} and \ref{scaling-agentsloglog}, we find a linear scaling of the Cross model with regard to the 
number of time steps and number of agents used in the simulation.
This is an expected result and seems to be a universal observation for our tool.
Although the example in figure \ref{scaling-timestepsWMatlabloglog} is conducted with the Cross model, we observe linear scaling for the Harras and LLS model with respect to time steps, as well.
Figure \ref{scaling-agentsLLS} reveals linear scaling of the LLS model with respect to the number of agents. 
For every data point we averaged the runtime of 100 simulations. We used the most basic setup relying on the standard C++ pseudo random number generator and using \texttt{.csv} files as an output. With respect to the runtime one might consider this choice as the  worst case. Using the pseudo random generator from the Intel MKL and output in \texttt{HDF5} files we achieve even faster runtimes. An example is given in \cref{SpeedUp}.

Finally, the generality of the meta-model presented in section 2, comes at the cost of preventing the exploitation of the parallelism and suitability for vectorization inherent to ABCEM models for which the decision processes of all agents can be calculated simultaneously. Since only very simple ABCEM models exhibit such inherent parallelism, this drawback is easily outweighed by the benefit of the ease of implementation and recombination of ABCEM models and the performance of simulations of these models.


\subsection{Pseudo Random Numbers}
\label{nca-rngs}
\label{RandomNumbers}
As mentioned before, many ABCEM models heavily utilize pseudo random numbers.
Hence, quality and efficiency of pseudo random number generators directly influence the quality of simulation results.
We start with a calculation example in order to quantify the number of pseudo random numbers possibly needed in our simulator.
Then, we discuss the implemented pseudo random number generators in SABCEMM.
Finally, we investigate two aspects related to the generation of pseudo random numbers: efficient generation of large amounts of pseudo random numbers and influences of different pseudo random number generators on simulation results. \\ \\
To stress the importance of efficiently generating large amounts of pseudo random numbers, we assume a simulation with $10,000$ time steps.
This provides a sufficiently large sample size for proper statistical analysis.
In addition, we assume that the market mechanism requires one pseudo random number per time step. Table \ref{tableRN} presents the number of pseudo random number needed for varying number of agents and different amounts of pseudo random numbers needed for each agent per time step.
From table \ref{tableRN}, we see that even for the small number of 100 agents we already need one million pseudo random numbers. 
\begin{table}
\begin{center}
  \begin{tabular}{| l || c | c |r|}
  \hline
  number of agents & 
  \multicolumn{3}{|c|}{number of random numbers}\\
   & 
  \multicolumn{3}{|c|}{per agent and time step}\\
    \hline
      & 1 & 2 &3\\ 
     \hline \hline
 100 & 1,010,000&2,010,000  & 3,010,000\\ \hline
1,000 &10,010,000  &20,010,000  &30,010,000\\ \hline
10,000 & 100,010,000 & 200,010,000 &300,010,000\\ \hline
  \end{tabular}
  \caption{Example calculation of needed random numbers.}\label{tableRN}
\end{center}
\end{table}

\paragraph{Pseudo Random Number Generators in SABCEMM}
 As the previous calculation reveals, ABCEM models may require a large amount of pseudo random numbers.
 The SABCEMM simulator supports multiple pseudo random number generators, namely the NAG library \cite{NAGlibrary}, the Intel Math Kernel Library (MKL) \cite{intelMKL} and the pseudo random generator of the C++ library.
The chosen pseudo random number generator in each library are variants of the Mersenne Twister pseudo random number generator \cite{matsumoto1998mersenne}.
More precisely, in the C++ library we have chosen \texttt{mt19937\_64}, in the Intel Math Kernel Library \texttt{VSL\_BRNG\_MT2203} and in the NAG library \texttt{g05dyc, g05dac, g05ddc}.
The testing of pseudo random number generators is a severe issue and has been first suggested by Knuth \cite{knuth1997art}.
The recently introduced test library TestU01 \cite{l2007testu01} demonstrates the goodness of the Mersenne Twister for large scale applications.\\ 
While the pseudo random generator offered by the C++ library has the advantage that it is shipped with every C++ compiler we advise strongly against any standard older then C++11.
The quality of the pseudo random numbers generated by older standards do not meet our requirements.
The Intel MKL library and the NAG library have to be provided by the user at compile time.
The version depends on what software is installed on your system.
We rely on the Intel\textcopyright Math Kernel Library 11.3.3 for Linux to provide our pseudo random numbers if not otherwise noted.
As shown in subsequent paragraph it is very fast when using the batch mode.
The SABCEMM simulator allows the user to choose the library best suited to his needs.

\paragraph{Efficient Generation of Pseudo Random Numbers}

\begin{figure}[h!]
    \begin{center}
        \includegraphics[width=0.7\textwidth]{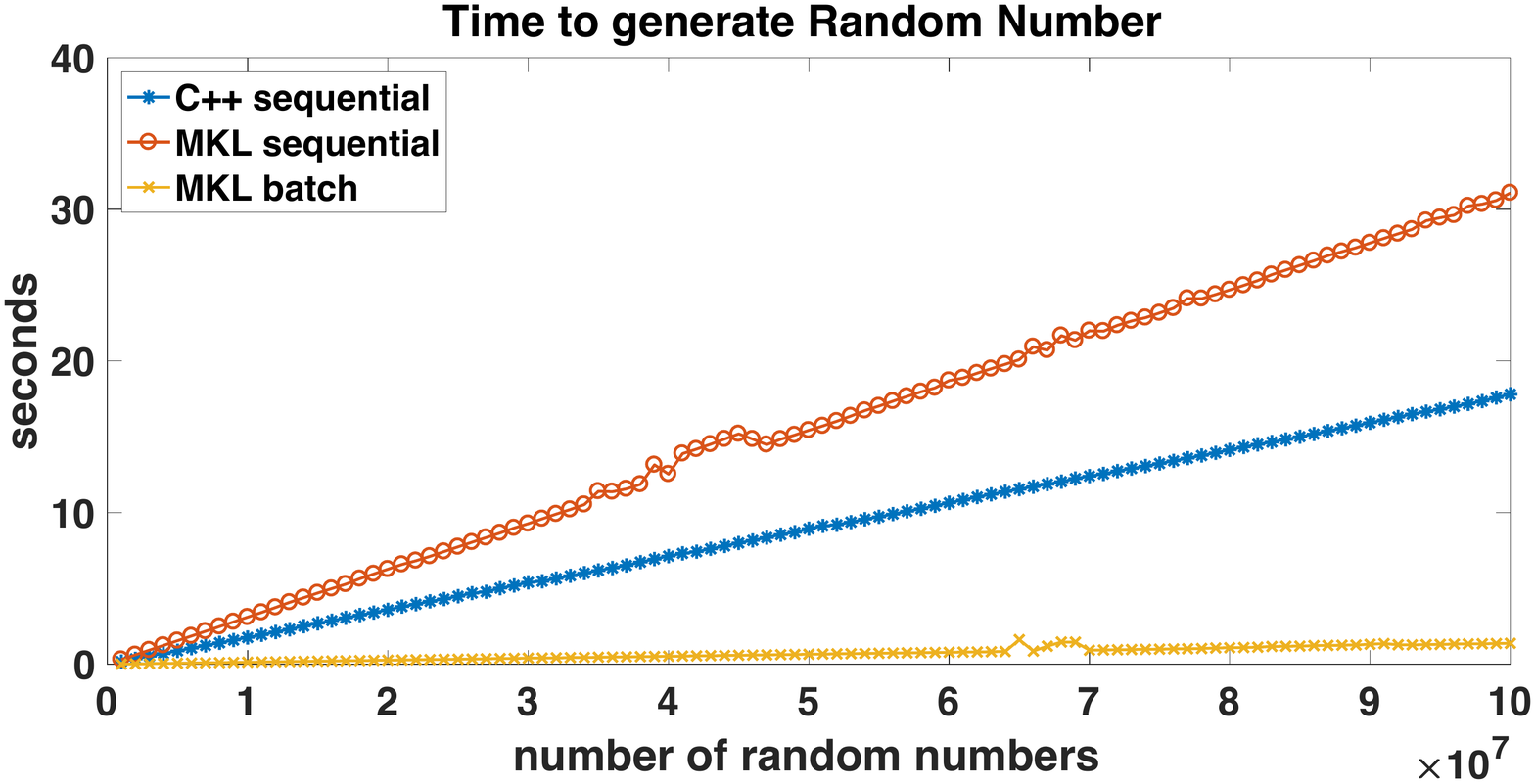}
        \caption{Time needed to generate a variety of pseudo random numbers.}
        \label{scaling-rngtimings}
    \end{center}
\end{figure}

In order to avoid the overhead implied by invoking the pseudo random number generator every time a pseudo random number is required during the simulation, we introduce the possibility to generate a pool of pseudo random numbers into the SABCEMM simulation software. Figure \ref{scaling-rngtimings} reveals the speed of each generator regarding the creation of different pool sizes.

Pseudo random numbers then can be drawn from this pool instead of being computed on the fly.
In addition, it is also possible to calculate pseudo random numbers on the fly the moment they are required.
Table \ref{RuntimeRN} summarizes the change of runtime of the Harras model with respect to the C++ sequential and MKL batch pseudo random number generator.  

\begin{table}[h!]
    \begin{center}
\begin{tabular}{|c|c||c|c|c|c|}
    \hline 
    \multicolumn{2}{|c||}{Amount of}   & \multicolumn{2}{c|}{Runtime for the simulations}  & \multicolumn{2}{c|}{Generating random numbers}\\ 
    \hline 
     Agents &  Random Numbers & C++ sequential & MKL batch & C++ sequential & MKL batch \\ 
    \hline 
    2,500 & 25,040,002 & 8.65 sec & 5.61 sec & 4.46 sec & 0.346 sec \\ 
    \hline 
    5,041 & 50,480,494 & 18.39 sec & 13.25 sec & 8.99 sec & 0.697 sec \\ 
    \hline 
    7,569 & 75,790,830 & 28.46 sec & 20.34 sec & 13.50 sec & 1.047 sec \\ 
    \hline 
    10,000 & 100,130,002 & 39.52 sec & 30.63 sec & 17.84 sec & 1.383 sec \\ 
    \hline 
\end{tabular} 
        \caption{Runtime of the Harras model with respect to varying number of agents and different pseudo random number generators. Further parameters are set to the values in table \ref{harras-basic-parameter}.}\label{RuntimeRN}
    \end{center}
\end{table}

From this, we can easily see that pooling of pseudo random numbers is well-suited to reduce the overall runtime of simulations carried out using the SABCEMM simulation software. 
Finally, table \ref{SpeedUp} shows considerable speed up by utilizing the MKL batch pseudo random number generator. We obtain a maximal speed up of $35\%$ of the total simulation time. \\ \\

\begin{table}[h!]
    \begin{center}
\begin{tabular}{|c|c||c|c|}
    \hline 
    \multicolumn{2}{|c||}{Amount of}   & \multicolumn{2}{|c|}{Speed up} \\ 
    \hline 
    Agents &  Random Numbers & of the simulations  & for generating random numbers \\ 
    \hline 
    2,500 & 25,040,002 & 35 \% & 92 \% \\ 
    \hline 
    5,041 & 50,480,494 & 29 \% & 92 \%  \\ 
    \hline 
    7,569 & 75,790,830 & 28 \% & 92 \%\\ 
    \hline 
    10,000 & 100,130,002 & 22 \% & 92 \% \\ 
    \hline 
\end{tabular} 
        \caption{Speed up of the Harras model due to the MKL batch pseudo random number generator.} \label{SpeedUp}
    \end{center}
\end{table}
Besides the efficient generation of pseudo random numbers, the quality of the generated pseudo random numbers itself are of paramount importance.  
It has been shown that linear congruential pseudo random number generators such as the \texttt{RANDU} generator have a poor performance in large scale applications \cite{hellekalek1998good, knuth1997art}. Unfortunately the \texttt{RANDU} generator has been widely used in the 1970s and is possibly still used until today. For that reason, SABCEMM does not support the \texttt{RANDU} generator.

\section{Conclusion}
\label{sec-conclusion}
We have introduced the large scale open-source simulator SABCEMM, especially designed for heterogeneous multi-agent ABCEM models.
Although there is a huge number of simulators for general agent-based models \cite{abar2017agent, allan2010survey, nikolai2009tools, tobias2004evaluation}, a simulator specialized to ABCEM models was missing. 
Nevertheless, we want to mention the small scale JAMEL and medium scale JASA simulators \cite{abar2017agent}. The former has been used to study macroeconomic models, whereas the latter considers general double-auction markets. In comparison to the previous examples the SABCEMM simulator is built for a very general ABCEM model as introduced in \cref{model}.\\ \\
In section \cref{model} we have introduced the abstract ABCEM model and the new concept of an environment. Furthermore, we have motivated the model from economic perspective and have shown the huge adaptability of the model. In section \ref{sec-software-architecture} we have presented the software architecture. Especially we have presented the object-orientation and how this functionality enables the user to build new ABCEM models. In addition, we discussed computational aspects of the SABCEMM simulator. Hence, we obtained a linear scaling of our simulator with respect to the number of time steps and agents. Furthermore, our examples indicate that a good and clever integration of a pseudo random number generator can reduce the run time up to $35\%$.
Finally, we aim to summarize the distinct features of the  SABCEMM simulator. 
\begin{enumerate}
	\item \textbf{Generality}: It is built on the basis of a very general ABCEM model, suitable for implementation of a very wide class of ABCEM models.
	\item \textbf{Recombination}: It allows recombination of the building blocks of different ABCEM models via configuration files for evaluation of novel ABCEM models.
	\item \textbf{Comparability}: It provides a common foundation, including random number generators, for fair comparisons of different ABCEM models.
	\item \textbf{Extensibility}: Implementation of additional ABCEM models is facilitated by object-oriented design.
	\item \textbf{Efficiency}: Suitability for simulations including very large, more than $10^6$ agents, allowing for testing for finite size effects.
\end{enumerate}
%
%
%

In a subsequent publication we will demonstrate the great flexibility of the SABCEMM simulator, for example, by interchanging the market mechanism and agent type.
Furthermore, we plan to discuss the simulation output of different models and aim to compare these results with regard to the reproduced stylized facts. Future work is intended to include the implementation of additional ABCEM models and the utilization of the SABCEMM simulator for computing model sensitivities and perform parameter fitting for ABCEM model.
%


\section*{Acknowledgement}
Torsten Trimborn greatfully acknowledges support by Hans-Böckler Stiftung.

\appendix

\section{Appendix}
\label{sec-appendix}

\subsection{Models}
\label{appendixModel}

\paragraph{Cross Model}
We present the Cross model as defined in \cite{cross2005threshold}.\\ \\
We assume a fixed number of $N\in\N$ agents. Each agent decides in each time step, whether he wants to be long or short in the market. Thus, the investment propensity $\sigma_i,\ 1\leq i\leq N$ of each agent switches between $\sigma_i=\pm 1$. The excess demand of all investors at time $[0,\infty)$ is then defined as:
\[
ED(t):= \frac{1}{N}\sum\limits_{i=1}^N\sigma_i(t).
\] 
Furthermore, the model introduces two pressures, the \textit{herding pressure} and the \textit{inaction pressure}, which control the switching mechanism.\\
The \textit{inaction pressure} is defined by the interval
\[
I_i=\left[ \frac{m_i}{1+\alpha_i}, m_i\ (1+\alpha_i)\right],
\] 
where $m_i$ denotes the stock price of the last switch of agent $i$ and $\alpha_i>0$ is the so called \textit{inaction threshold}.
The \textit{herding pressure} is given by:
\[ \begin{cases} c_i(t+\Delta t)= c_i(t)+ \Delta t |ED(t)|,& \text{if}\ \sigma_i(t)\ ED(t)<0\\
			c_i(t+\Delta t)=c_i(t),& \text{otherwise}.
			 \end{cases}
			\]
The implementations of herding pressure and inaction pressure can be found in the Git repository\footnote{Code excerpt: \url{https://github.com/SABCEMM/SABCEMM/blob/v0.1-alpha/src/Agent/AgentCross.cpp\#L138-L140}} \footnote{Code excerpt: \url{https://github.com/SABCEMM/SABCEMM/blob/v0.1-alpha/src/Agent/AgentCross.cpp\#L129-L133}}, respectively.

In addition, one defines the  \textit{herding threshold} $\beta_i$. The thresholds are chosen once randomly from an i.i.d. random variable, which is uniformly distributed. 
\begin{align*}
\alpha_i\sim \Unifc(A_1,A_2), \ A_2>A_1>0,\\
\beta_i\sim \Unifc(B_1,B_2), \ B_2>B_1>0.
\end{align*}

The constants $B_1$ and $B_2$ have to scale with time, since they correspond to the time units an investor can resist the herding pressure.
\begin{align*}
& B_1:= b_1\cdot \Delta t,\\
&B_2:= b_2\cdot \Delta t.
\end{align*}
\paragraph{Switching mechanism}
	The switching is then induced if 
	\[
	c_i>\beta_i\ \text{or}\  S(t)\notin I_i.
	\]
After a switch the herding pressure is reset to zero and the inaction interval gets updated as well. The stock price is then driven by the excess demand:
\begin{align*}
&S(t+\Delta t)=S(t)\ \exp\{  (1+\theta\ |ED(t)| )\ \sqrt{\Delta t}\ \eta(t)+\kappa\ \Delta t\ \frac{\Delta ED(t)}{\Delta t} \},\\
&\sqrt{\Delta t}\ \eta\sim \mathcal{N}(0,\Delta t)\\
&\Delta ED(t):=\frac{1}{N}\sum\limits_{i=1}^N \sigma_i(t)-\frac{1}{N}\sum\limits_{i=1}^N \sigma_i(t-\Delta t),
\end{align*}
where $\kappa$  denotes the market depth and $\Delta t >0$ the time step.\\

In SABCEMM, the price evolution\footnote{Code excerp: \url{https://github.com/SABCEMM/SABCEMM/blob/v0.1-alpha/src/PriceCalculator/PriceCalculatorGeneral.cpp\#L92}} is implemented as $$ S(t+\Delta t) = S(t) + \Delta t*F+sqrt(\Delta t) * G * \eta $$  with suitable $F$ \footnote{Code excerpt: \url{https://github.com/SABCEMM/SABCEMM/blob/v0.1-alpha/src/PriceCalculator/PriceCalculatorGeneral.cpp\#L173-L17}} and $G$\footnote{Code excerpt: \url{https://github.com/SABCEMM/SABCEMM/blob/v0.1-alpha/src/PriceCalculator/PriceCalculatorGeneral.cpp\#L190-L193}}.

\paragraph{Cross model extensions:}
One alternative pricing function is given by:
\begin{align*}
S(t+\Delta t)=S(t)\ + \Delta t\  \kappa\ \frac{\Delta ED(t)}{\Delta t} \ S(t)+ \sqrt{\Delta t}\ F_{Cross}(S,ED)\ S(t)\  \eta,\\
\end{align*}
Furthermore, we have added the wealth evolution, for a fixed interest rate $r>0$ and fixed investment fraction $\gamma\in (0,1)$:
$$
w_{i}(t+\Delta t) = w_i(t)+ \Delta t\ \left[(1-\gamma)\ r + \gamma \frac{S(t)-S(t-\Delta t)}{\Delta t\ S(t)}\right] w_i(t).
$$

\paragraph{LLS Model}
We have implemented the model as defined in \cite{levy1994microscopic, levy1995microscopic}. 
In comparison to the original model, we introduce one possible time scaling. In order to obtain the original model one needs to set $\Delta t=1$.\\ \\
The model considers $N\in\N$ financial agents who can invest $\gamma_i\in [0.01, 0.99],\ i=1,...,N$ of their wealth $w_i\in\R_{>0}$ in a stocks and have to invest $1-\gamma_i$ of their wealth in a safe bond with interest rate $r\in(0,1)$. The investment propensities $\gamma_i$ are determined by a utility maximization and the wealth dynamic of each agent at time $t\in[0,\infty )$ is given by
\begin{footnotesize}
\begin{align*}
&{w}_i(t)=w_i(t-\Delta t)\\
&\quad +\Delta t \left((1-\gamma_i(t-\Delta t))\ r\ w_i(t-\Delta t)+\gamma_i(t-\Delta t)\ w_i(t-\Delta t)\ \underbrace{\frac{\frac{{S}(t)-S(t-\Delta t)}{\Delta t}+D(t)}{S(t)}}_{=:x(S,t,D)}\right).
\end{align*} 
 \end{footnotesize}
The dynamics is driven by a multiplicative dividend process. Given by:
\[
D(t):=(1+\Delta t\  \tilde{z})\  D(t-\Delta t),
\]
 where $\tilde{z}$ is a uniformly distributed random variable with support $[z_1,z_2]$.  
The price is fixed by the so called \textit{market clearance condition}, where $n\in\N$ is the fixed number of stocks and $n_i(t)$ the number of stocks of each agent.  
\begin{align}
n=\sum\limits_{i=1}^N n_i(t)=\sum\limits_{k=1}^N \frac{\gamma_k(t)\ w_k(t)}{S(t)}. \label{fixedpointLLS}
\end{align}
The utility maximization is given by
\[
\max\limits_{\gamma_i \in [0.01,0.99]} E[\log(w(t+\Delta t,\gamma_i,S^h))].
\]
with
\begin{align*}
E[\log(w(t+\Delta t, \gamma_i,S^h))]=\frac{1}{m_i} \sum\limits_{j=1}^{m_i} U_i\Bigg(&(1-\gamma_{i}(t)) w_{i}(t,S^h) \left(1+r\Delta t\right)\\ 
&+\gamma_{i}(t) w_{i}(t, S^h) \Big(1+x\big(S,t-j\Delta t,D\big)\ \Delta t\Big)\Bigg).
\end{align*}

The constant $m_i$ denotes the number of time steps each agent looks back. Thus, the number of time steps $m_i$ and the length of the time step $\Delta t$ defines the time period each agent extrapolates the past values. The superscript $h$ indicates, that the stock price is uncertain and needs to be fixed by the market clearance condition.
Finally, the computed optimal investment proportion gets blurred by a noise term.
\[
\gamma_i (t)=\gamma_i^{*}(t)+\epsilon_i,
\]
where $\epsilon_i$ is distributed like a truncated normally distributed random variable with standard deviation $\sigma_{\gamma}$.

\paragraph{Utility maximization}
Thanks to the simple utility function and linear dynamics we can compute the optimal investment proportion in the cases where the maximum is reached at the boundaries.
The first order necessary condition is given by:
$$
f(\gamma_i):= \frac{d}{dt} E[\log(w(t+\Delta t, \gamma_i,S^h))] = \frac{1}{m_i} \sum\limits_{j=1}^{m_i} \frac{\Delta t\ (x\big(S,t-j\Delta t,D\big)-r)}{\Delta t\ (x\big(S,t-j\Delta t,D\big)-r)\ \gamma_i+ 1+\Delta t\ r}.
$$
 Thus, for $f(0.01)<0$ we can conclude that $\gamma_i=0.01$ holds. In the same manner, we get $\gamma_i=0.99$, if $f(0.01)>0$ and $f(0.99)>0$ holds. 
 Hence, solutions in the interior of $[0.01, 0.99]$ can be only expected in the case: $f(0.01)>0$ and  $f(0.99)<0$. This coincides with the observations in \cite{samanidou2007agent}.

\paragraph{Harras model}
\label{appendix-harras-model}
We present the Harras model as defined in \cite{harras2011grow}.\\ \\
We consider $N$ financial agents where each agent is equipped with a personal opinion $\psi_i(t_k)$, and $t_k$ denotes a discrete time step. The personal opinion is created through the personal information of each agent $\epsilon_i(t_k)$, public information $n(t_k)$ and the expected action of the surrounded neighbor $j$ by the agent $i$, $E_i[\sigma_j(t_k)], \sigma_j\in\{-1,0, 1\}$. The opinion of the i-th agent at time $t_k$ reads:
\begin{align}
\psi_i(t_k) = c_{1,i} \sum\limits
_{j=1}^4  k_{ij}(t_{k-1})\ E[\sigma_i(t_k)]   + c_{2,i} u(t_{k-1})\ n(t_k) +c_{3,i}\ \epsilon_i(t_k). \label{opinion}
\end{align}
During the evaluation of our simulations we noticed a significant difference in the magnitude of the price's volatility. Our investigation leads us to the conclusion that the opinion of the i-th agent at time $t_k$ should instead be: 
\[
\psi_i(t_k) = c_{1,i} \frac{1}{4} \sum\limits
_{j=1}^4  k_{ij}(t_{k-1})\ E[\sigma_i(t_k)]   + c_{2,i} u(t_{k-1})\ n(t_k) +c_{3,i}\ \epsilon_i(t_k). 
\]
The weights $(c_{1,i}, c_{2,i}, c_{3,i})$ are chosen initially for each agent from three uniformly distributed random variables on the domains $[0,C_l],\ l\in \{1, 2, 3 \}$.
The private and public information $\epsilon_i(t_k), n(t_k)$ are modeled as standard normally distributed i.i.d. random variables. 
The agents are grouped on a virtual square lattice with periodic boundary conditions, such that each agent has four neighbors. 
We update of the opinion of each agent is performed in each time step in random order. 
The additional factor $k_{ij}$ weights the predicted action of the j-th agent based on the past performance. In the same manner the factor $u$ weights the public information stream. The update rule of these weighting factors is given by:
\begin{align*}
&u(t_k) = \alpha\ u(t_{k-1}) +(1-\alpha)\ n(t_{k-1}) \frac{ED(t_k)}{\sigma_{ED}(t_k)},\\
&k_{i,j} (t_k)= \alpha\ k_{i,j} (t_{k-1})+(1-\alpha)\ E_i[\sigma_j(t_{k-1})]  \frac{ED(t_k)}{\sigma_{ED}(t_k)}, 
\end{align*}
with the constant $0<\alpha<1$ and the volatility
\begin{align*}
&\sigma_{ED}^2(t_k)= \alpha\ \sigma_{ED}^2(t_{k-1})+(1-\alpha)\ (ED(t_{k-1})- \langle ED(t_k)\rangle)^2,\\
&\langle ED(t_k)\rangle = \alpha\ \langle ED(t_{k-1}) \rangle +(1-\alpha)\ ED(t_{k-1}), 
\end{align*}
where the brackets $\langle\cdot \rangle $ denote the expected excess demand $ED$. The agent's action on the market is then determined 
by a threshold of each agent. The threshold $\bar{\psi}_i$ is drawn from a uniform distribution in the interval $[0,\Omega]$. The trading decision of each agent is characterized by $\sigma_i=\pm 1$, where $\sigma_i =1$ represents a buy order and $\sigma=-1$ a sell order.  We have
\begin{align*}
\sigma_i(t_k)=
\begin{cases}
& 1,\quad\psi_i(t_k)> \bar{\psi}_i,\\
&-1,\quad  \psi_i(t_k)< -\bar{\psi}_i,\\
&0,\quad \text{else}.
\end{cases}
\end{align*}
Furthermore, each agent tracks his number of stocks $q_i$ and the cash $w_i$ and thus the trading volume $v_i(t_k)$ of each agent (in units of stocks) is given by
\begin{align*}
v_i(t_k)= 
\begin{cases}
g\ \frac{w_i(t_k)}{S(t_{k})},\quad \sigma_i=1,\\
g\ q_i(t_k),\quad \sigma=-1.
\end{cases}
\end{align*}
Here, $S(t_k)$ denotes the stock price and $g\in (0,1)$ a fixed fraction of wealth each agent wants to invest in stocks. The stock price is then driven by the
excess demand $ED(t_k)$
$$
\log(S(t_k)) = \log(S(t_{k-1}))+ ED(t_{k-1}),
$$ 
where $ED$ is defined as:
$$
ED(t_k):= \frac{1}{\lambda\ N} \sum\limits_{i=1}^N \sigma_i(t_{k})\ v_i(t_{k}).
$$
The constant $\lambda>0$ represents the market depth. Finally, we want to state the update mechanism of $w_i$ and $q_i$. 
\begin{align*}
&w_i (t_k)= w_i(t_{k-1}) -\sigma_i(t_{k-1})\ v_i(t_{k-1})\ S(t_k),\\
&q_i(t_k) = q_i(t_{k-1})+\sigma_i(t_{k-1})\ v_i(t_{k-1}).
\end{align*}

\subsection{Technical Details}
\label{technicalDetails}
\begin{itemize}
\item \textbf{Numerical Discretization}: With numerical discretization we refer to approximation concepts for time continuous ODEs. The most prominent numerical schemes are the explicit and implicit Euler discretization. We refer to \cite{butcher2016numerical} for a detailed discussion. 
\item \textbf{Wiener process}: A Wiener process is time continuous stochastic process and plays a prominent role in the definition of a stochastic integral. We refer to \cite{evans2012introduction} for details.
\item \textbf{It\^o SDE}: The term It\^o SDE refers to a SDE defined by the It\^o stochastic integrals. For details we refer to \cite{evans2012introduction}.
\item \textbf{Stratonovich SDE}: The term Stratonovich SDE refers to a SDE defined by the Stratanovich stochastic integrals. For details we refer to \cite{evans2012introduction}.
\end{itemize}

\FloatBarrier

\subsection{Parameter sets}
\paragraph{Cross Model}

For simulations using the Cross model, we use the parameters and initial values presented in table \ref{cross-basic-parameter}.
\FloatBarrier
\begin{table}[h!]
	\begin{subtable}[b]{0.45\textwidth}
	    \begin{center}
	        \begin{tabular}{|c||c|}
	            \hline
	            Parameter & Value\\
	            \hline
	            \hline
	            $N$ & $1000$ \\
	            \hline
	            $A_1$& $0.1$\\
	            \hline 
	            $A_2$ & $0.3$\\
	            \hline
	            $b_1$& $25$\\
	            \hline 
	            $b_2$ & $100$\\
	            \hline 
	            $w_i(t=0)$ & $1 \quad \forall 1\leq i\leq N$\\
	            \hline 
	            time steps & $10,000$ \\
	            \hline
	            $\Delta t$ &$ 4\cdot 10^{-5} $\\
	            \hline
	            $\kappa$ & $0.2$\\
	            \hline
	            $\theta$& $0$\\
	            \hline
	            $S(t=0)$ & $1$\\
	            \hline 
	        \end{tabular} 
		\end{center}
		\caption{Parameters of Cross model. }
	\end{subtable}
	\hspace{0.2cm}
	\begin{subtable}[b]{0.45\textwidth}
		\begin{center}
	        \begin{tabular}{|c||c|}
	            \hline
	            Variable & Initial Value\\
	            \hline
	            \hline
	            $ED(t=0)$& $\frac{1}{N}\sum\limits_{i=1}^N \gamma_i(0)$\\
	            \hline
	            $c_i(t=0)$ & $B_1+\texttt{rand}\ (B_2-B_1),\ \forall 1\leq i\leq N$\\
	            \hline
	            $m_i(t=0)$ & $S(t=0),\ \forall 1\leq i\leq N$\\
	            \hline
	            $\sigma_i(t=0)$ & $ \Unifd(\{-1,1\})$ \\
	            \hline
	        \end{tabular}
	    \end{center}
	    \caption{Initial values of Cross model.}
	\end{subtable}
	\caption{Cross basic setting.}
    \label{cross-basic-parameter}
\end{table}
\FloatBarrier

\paragraph{LLS Model}
The initialization of the stock return is performed by creating an artificial history of stock returns.
The artificial history is modeled as a Gaussian random variable with mean $\mu_h$ and standard deviation $\sigma_h$.
Furthermore, we have to point out that the increments of the dividend is deterministic, if $z_1=z_2$ holds.
We used the C++  standard pseudo random number generator for all simulations of the LLS model if not otherwise stated.
The parameters and initial values used for simulations using the LLS model are shown in tables \ref{LLS-basic} and \ref{LLS-3-agents}.

\FloatBarrier
\begin{table}
	\begin{subtable}[b]{0.45\textwidth}
		\begin{center}
			\begin{tabular}{|c||c|}
			\hline
			Parameter & Value\\
			\hline
			\hline
			$N$ & $100$\\
			\hline
			$m_i$& $15$\\
			\hline 
			$\sigma_{\gamma}$ & $ 0 $ or $0.2$\\
			\hline
			$r$& $0.04$ \\
			\hline 
			$z_1=z_2$& $0.05$\\
			\hline
			$\Delta t$ &$ 1$\\
			\hline
			time steps & 200 \\
			\hline
			\end{tabular}
		\end{center}
		\caption{Parameters of LLS model.}
	\end{subtable}
	\hspace{0.5cm}
	\begin{subtable}[b]{0.45\textwidth}
		\begin{center}
			\begin{tabular}{|c||c|}
			\hline
			 Variable & Initial Value\\
			\hline
			\hline
			$\mu_h$ & $0.0415$\\
			\hline
			$\sigma_h$ & $0.003$\\
			\hline
			$\gamma(t=0)$ & $0.4$\\
			\hline
			 $w_i(t=0)$ & $1000$ \\
			 \hline
			 $n_i (t=0)$ & $100$ \\
			 \hline
			 $S (t=0)$ & $4$ \\
			 \hline
			 $D (t=0)$ & $0.2$ \\
			 \hline
			\end{tabular}
		\end{center}
		\caption{Initial values of LLS model.}
	\end{subtable}
	\caption{Basic setting of the LLS model.} \label{LLS-basic}
\end{table}

\begin{table}
	\begin{subtable}[b]{0.45\textwidth}
		\begin{center}
			\begin{tabular}{|c||c|}
			\hline
			Parameter & Value\\
			\hline
			\hline
			$N$ & $99$\\
			\hline
			$m_i $&  $10,\ 1 \leqslant i \leqslant 33$  \\
			                  & $141,\ 34 \leqslant i \leqslant 66$ \\
			                  & $256,\ 67 \leqslant i \leqslant 99$ \\
			\hline 
			$\sigma_{\gamma}$ & $ 0.2 $\\
			\hline
			$r$& $0.0001$ \\
			\hline 
			$z_1=z_2$& $0.00015$\\
			\hline
			$\Delta t$ &$ 1 $\\
			\hline 
			time steps & $20,000$\\
			\hline
			\end{tabular}
		\end{center}
		\caption{Parameters of LLS model.} 
	\end{subtable}
	\hfill
	\begin{subtable}[b]{0.45\textwidth}
		\begin{center}
			\begin{tabular}{|c||c|}
			\hline
			 Variable & Initial Value\\
			\hline
			\hline
			  $\mu_h$ & $0.0415$ \\
			  \hline
			  $\sigma_h$ &$ 0.003$ \\
			  \hline
			  $\gamma_i(t=0)$ & $0.4$ \\
			  \hline
			  $w_i (t=0)$ & $1000$ \\
			  \hline
			   $n_i (t=0)$ & $100$ \\
			   \hline
			    $S (t=0)$ & $4 $\\
			    \hline
			   $D (t=0)$ & $0.004$ \\
			 \hline
			\end{tabular}
		\end{center}
		\caption{Initial values of LLS model.}
	\end{subtable}
	\caption{Setting for the LLS model (3 agent groups).}
	\label{LLS-3-agents}
\end{table}
\FloatBarrier

\paragraph{Harras Model}
The parameters and initial values used for simulations using the Harras Model are shown in table \ref{harras-basic-parameter}.

\FloatBarrier
\begin{table}[h!]
	\begin{subtable}[b]{0.45\textwidth}
    	\begin{center}
        	\begin{tabular}{|c||c|}
	            \hline
	            Parameter & Value\\
	            \hline
	            \hline
	            $C_1$ & $0$\\
	            \hline
	            $C_2$ & $1$\\
	            \hline 
	            $C_3$ & $1$\\
	            \hline
	            $\Omega$& $2$\\
	            \hline 
	            $g$ & $0.02$\\
	            \hline
	            $\alpha$& $0.95$\\
	            \hline
	            $w_i(t_0)$ & $1 \quad \forall 1\leq i\leq N$\\
	            \hline 
	            $q_i(t_0)$ & $1 \quad \forall 1\leq i\leq N$\\
	            \hline
	            $N$ & $2500$\\
	            \hline  
	            $\lambda$ & $0.25$\\
	            \hline
	            $S(t=0)$ & $1$\\
	            \hline 
	            $q_i(t=0)$ & $1 \quad \forall 1\leq i\leq N$\\
	            \hline 
	            time steps & $10,000$\\
	            \hline 
        	\end{tabular}
        \end{center}
        \caption{Parameters in the Harras model.}
	\end{subtable}
	\hfill
	\begin{subtable}[b]{0.45\textwidth}
		\begin{center}
	        \begin{tabular}{|c||c|}
	            \hline
	            Variable & Initial Value\\
	            \hline
	            \hline
	            $k_{ij}(t=0)$& $\Unifc(0,1)$\\
	            \hline
	            $E_i[\sigma_j(t=0)]$ & $\Unifd(\{-1,0,1\})$\\
	            \hline
	            $\sigma_{ED}^2(t=0)$ & $0.1$\\
	            \hline
	            $ \langle ED(t=0)\rangle$ & $0$\\
	            \hline
	            $ED(t=0)$ & $0$\\
	            \hline
	            $ED(t={-1})$ & $0$\\
	            \hline
	            $u_i(t={-1})$ & $0$\\
	            \hline
	            $u_i(t=0)$ & $0$\\
	            \hline
	            $v_i(t=0)$ & $0$\\
	            \hline
	            $\sigma_i(t=0)$ & $0$\\
	            \hline
	        \end{tabular}
	    \end{center}
	    \caption{Initial values Harras.}
	\end{subtable}
	\caption{Harras basic setting.}
    \label{harras-basic-parameter}
\end{table}

\clearpage
\subsection{Software}

We aim to explain how we translate our abstract meta-model into software instructions. 
The centerpiece of our software is the calculation of the excess demand and the price calculator in each time step.
Furthermore, one needs to update the action of each agent, which will depend on the stock price and probably on the excess demand. 
\begin{lstlisting}[frame=single]
excessDemandCalculator = ExcessDemandCalculator(agents, input)
priceCalculator = PriceCalculator(excessDemandCalculator)

for t in [0,N]:
    excessDemand = excessDemandCalculator.timestep(t)
    price = priceCalculator.timestep(excessDemand, t)
    
    for agent in agents:
        agent.timestep(excessDemand, price, agents)

\end{lstlisting}

\paragraph{Software Classification}
\label{software-classification}
To facilitate the selection of a proper software tool for other researchers, we classify SABCEMM by the five categories of Nikolai and Madey \cite{nikolai2009}:
\begin{itemize}
    \item \textbf{ Programming language required to create a model or simulation}
    \begin{itemize}
        \item[] To introduce a new building block programming in C++ is required. From existing building blocks new models/simulations can be configured with XML input files.
    \end{itemize}

    \item \textbf{Operating system required to run the toolkit}
\begin{itemize}
    \item[] SABCEMM runs on POSIX (e.g. Linux, macOS) systems with a recent C++ compiler (e.g. g++-7 \footnote{\url{https://gcc.gnu.org}}) and CMake\footnote{min. version 2.8.12; \url{https://cmake.org}} support.
\end{itemize}

    \item \textbf{Type of license governing the platform}
\begin{itemize}
    \item[] We distribute our software under the 3-clause BSD license\footnote{\url{https://opensource.org/licenses/BSD-3-Clause}}. Accordingly it classifies as free and open-source.
\end{itemize}

    \item \textbf{Primary domain for which the toolkit is intended}
\begin{itemize}
    \item[] Our software is designed for financial market simulations, i.e. it falls in the category Computational economics/Auction mechanisms. SABCEMM enables researchers to do large scale simulations with heterogeneous agent types.
\end{itemize}

    \item \textbf{Degree of support available to the user of the toolkit}
\begin{itemize}
    \item[] A \texttt{Reference Manual}\footnote{\url{https://sabcemm.github.io/SABCEMM/}}, a \texttt{User Guide}\footnote{\url{https://github.com/SABCEMM/SABCEMM/wiki/User-Guide}} and further documentation is provided. Further, bugs and questions can be submitted on GitHub.
\end{itemize}
\end{itemize}

	
		\bibliographystyle{abbrv}	
	\bibliography{Quellen/SABCEMM.bib}

\end{document}